\documentclass[fleqn,usenatbib]{mnras}
\usepackage{epsfig}
\usepackage{graphicx}
 \usepackage{multirow}
 \usepackage{array}
\usepackage{color}
\usepackage{amsmath}
\usepackage{subcaption}
\usepackage{txfonts}

\usepackage{amssymb}
\title{Theoretical investigations of propyl-cyanide formation in gas phase and on ice mantles}
\author[Kerkeni et al.]{Bouthe\"{i}na Kerkeni$^{1,2,3}$\thanks{E-mail:
boutheina.kerkeni@obspm.fr}, 
\newauthor Victoria G\'{a}mez$^4$, 
Ghofrane Ouerfelli$^{2,5}$, 
Maria Luisa Senent$^{4}$ 
\newauthor and Nicole Feautrier$^{3}$ \\ 
$^{1}$Institut Sup\'erieur des Arts Multim\'edia de la Manouba, Universit\'e de la Manouba, 2010 la Manouba, Tunisia\\
$^{2}$Facult\'e des Sciences de Tunis, D\'epartement de Physique, (LPMC), Universit\'e de Tunis El Manar, 2092 Tunis, Tunisia\\
$^{3}$Sorbonne Universit\'e, Observatoire de Paris, Universit\'e PSL, CNRS, LERMA, F-92195 Meudon, France \\
$^{4}$Departamento de Qu\'{\i}mica y F\'{\i}sica Te\'oricas, IEM-CSIC, Serrano 121, Madrid 28006, Spain \\
$^{5}$Department of Physics, College of Khurma University, Taif  University, P.O. Box11099, Taif 21944, Saudi Arabia
}

\date{Accepted XXX. Received YYY; in original form ZZZ}

\pubyear{2023}

\begin{document}
\label{firstpage}
\pagerange{\pageref{firstpage}--\pageref{lastpage}}

\maketitle

\begin{abstract}
Propyl cyanide (PrCN) (C$_3$H$_7$CN) with both linear and branched isomers is ubiquitous in interstellar space and is important for astrochemistry as it is one of the most complex molecules found to date in the interstellar medium. Furthermore, it is the only one observed species to share the branched atomic backbone of amino acids, some of the building blocks of life.  
Radical-radical chemical reactions are examined in detail using density functional theory, second order M$\phi$ller Plesset perturbation theory, coupled cluster methods, and the energy resolved master equation formalism to compute the rate constants in the low pressure limit prevalent in the ISM.
Quantum chemical studies are reported for the formation of propyl-cyanide (n-PrCN) and its branched isomer (iso-PrCN) from the gas phase association and surface reactions of radicals on a 34-water model ice cluster. We identify two and three paths for the formation of iso-PrCN, and n-PrCN respectively. The reaction mechanism involves the following radicals association: CH$_3$CHCH$_3$+CN, CH$_3$+CH$_3$CHCN for iso-PrCN formation and  CH$_3$CH$_2$+CH$_2$CN, CH$_3$+CH$_2$CH$_2$CN, CN+CH$_3$CH$_2$CH$_2$ leading to n-PrCN formation. We employ the M062X/6-311$++$G(d,p) DFT functional and MP2/aug-cc-pVTZ for reactions on the ice model, and gas phase respectively to optimize the structures, compute minimum energy paths and zero-point vibrational energies of all reaction mechanisms. In gas phase, the energetics of the five reactions are also calculated using the explicitly correlated cluster {\it ab initio} methods (CCSD(T)-F12). All reaction paths are exoergic and barrier-less in gas phase and on the ice-model suggesting that the formation of iso-PrCN and n-PrCN is efficient on the water-ice model adopted in this paper. The gas phase formation of iso-PrCN and n-PrCN however requires a third body or spontaneous emission of a photon in order to stabilize the molecules.

 \end{abstract}


\section{Introduction}

More than 200 species have been detected in the interstellar medium (ISM), among them many molecules, radicals and ions, containing the $-$C$\equiv$N functional group \citep{Solomon:1971,Mackellar:40,Guelin:1991,Kaiser:02,Cernicharo:2004,Margules:17, Belloche:14}. About 70$\%$ are organic in nature such as alcohols, aldehydes, acids, ethers, and amines. In fact, the strong bond of the cyano group $-$C$\equiv$N is present in more than 30 interstellar molecules \citep{Agundez:2013}. It is currently well accepted that these molecules play a major role in the chemical evolution of the interstellar and circumstellar media \citep{Herbst:95,Herbst:01}. The simplest cyanides HCN and H$_3$CN were detected since 1971 \citep{Snyder:1971,Solomon:1971}. After this discovery, many Complex Organic Molecules (COMs) such as CH$_2$CN  \citep{Irvine:1988}, CH$_3$CN \citep{Solomon:1971}, CH$_3$CHCN, and CH$_3$CH$_2$CN \citep{Johnson:1977} were observed in the ISM and were considered together with some other cyanopolyynes to play a crucial role in the chemical evolution of molecular clouds. Other studies \citep{Belloche:14} suggested that the growth of small cyanides towards largest ones is an important mechanism in the ISM chemistry. \\

The mechanism of formation of many complex organic molecules detected in star-forming regions of the interstellar medium, such as cold dense molecular clouds (T$\sim$ 10-100 K) and Hot Core (T$\sim$ 100-300 K) is not fully known \citep{Garrod:13,Garrod:2022}. In particular, the largest COM containing a cyano group is propyl cyanide (C$_3$H$_7$CN) which has the particularity of being the only one observed molecule in the ISM with a branched isomer (hereafter iso-PrCN). More recently, chiral although not branched molecules like propylene oxyde \citep{McGuire:2016}  and iso-propanol \citep{Belloche:2022} were detected in the same molecular cloud as PrCN isomers. Propyl cyanide has received increasing attention in recent years. Linear propyl cyanide (hereafter n-PrCN) was first observed toward the Galactic Center star-forming source Sagittarius B2(N) (hereafter Sgr B2(N)) \citep{Belloche:09}. The most stable isomer (iso-PrCN) was observed in 2014 with ALMA toward Sgr B2(N), with a fractional abundance with respect to H$_2$ of 1.3 $\pm$ 0.2 $\times$ 10$^{-8}$, leading to an iso/n abundance ratio of 0.40 \citep{Belloche:14}. This detection of a branched species was particularly important as iso-PrCN can be considered a precursor for branched amino acids with consequential implications towards complex species of prebiotic interest. Recently, the two isomers were detected with ALMA towards several emission peaks in Orion-KL \citep{Pagani:17} with different iso/n abundance ratios. 

This large saturated COM is not easily formed in the gas phase at temperatures below 100 K and the current mechanisms proposed to explain the gas phase abundance of interstellar PrCN are based on grain mantle chemistry. The formation mechanism of n-PrCN, was first investigated by \citep{Belloche:09} using a model proposed by  \citep{Garrod:2008}. They suggested that sequential addition of CH$_2$ and CH$_3$ radicals to CN, CH$_2$CN and C$_2$H$_4$CN on the ice mantle or direct addition of large hydrocarbon radicals such as C$_2$H$_5$ or C$_3$H$_7$ to CN is the fastest route to n-PrCN formation. They derived the column density of n-PrCN to be nearly 1.5 $\times$ 10$^{16}$ cm$^{-2}$ with a local cloud temperature of nearly 150 K.\\
Later, after the detection of iso-PrCN, the same group developed a new chemical kinetic model (MAGICKAL) in order to simulate the time-dependant chemistry of Sgr B2 \citep{Belloche:09,Garrod:13}. The model focused on the accretion and addition of small carbonaceous radicals at the ice mantle to pre-existing cyanides. However, their tool was not able to explain the observed abundance ratio iso/n. 

Recently, Garrod et al. \citep{Garrod:17} have studied the chemistry of Sgr B2(N) by considering the case of propyl and butyl cyanide using the same chemical model of  \citep{Belloche:14} with an updated chemical network. In this
model, they successfully obtained the detected abundance ratio between iso and n-propyl cyanide.

This suggests that an updated reaction network of a large and accurate set of reactions and rate coefficients in cloud modelling is timely. In particular, there is a need to accurately determine the efficiency of possible chemical reaction mechanisms leading to PrCN isomers formation, their energetic paths and chemical rate constants on the ice. \

In a previous work, we have investigated using highly correlated {\it ab initio} methods the gas phase bimolecular reaction of HCN with propene to produce the different C$_3$H$_7$CN isomers. However, although the process occurs through a concerted mechanism (C–C bond formation and proton transfer), due to the presence of a significant barrier, this reaction lead to very small rate constants \citep{Kerkeni:19}. Similar conclusion was found for the same reaction on an icy grain model \citep{Kerkeni:20}. Recently, \citep{Singh:2020} have proposed reactions in both the gas phase and interstellar ice of small cyanides with hydrocarbon radicals to explore the formation of either iso-PrCN and n-PrCN or ethyl and vinyl cyanide. It was found that the formation of PrCN from ethyl or vinyl cyanide is inefficient in the ISM dark clouds due to large entrance barriers in gas phase as well as in the ice.\\


One can imagine different reaction pathways, but most probable are those involving abundant, commonly found organic species. Some of such species are CN, CH$_3$, CH$_2$CN, C$_3$H$_7$ and C$_2$H$_5$. 
In this article, we focus on gas-phase and a solid-state chemistry formation processes where the ice model plays the role of the third body capable of stabilizing the formed iso-PrCN and n-PrCN molecules. We perform new electronic structure calculations on the energetics of this mechanism of approaching gas phase radicals on the ice mantle where they may get accreted. The chemical network consists of the radical-radical addition reactions upon the ice mantle. We build a new 34-waters amorphous ice model, and employ Density Functional Theory (DFT) for reactions on the ice model, the second order M$\phi$ller Plesset perturbation theory (MP2) and CCSD(T)-F12 for the quantum chemistry calculations in gas phase. Our main purpose is to provide accurate energetic data regarding adsorption energies and reaction paths of the proposed reactions. Variable reaction coordinate transition-state theory (VRC-TST) was used for rate constant calculations as we only have barrier-less reactions. 
The Rice-Ramsberger-Kassel-Marcus/Master-Equation (RRKM/ME) theory was used to calculate the low pressure limiting recombination rate constants of these channels as low pressure regime is prevailing in the interstellar medium. 
 The kinetics of the proposed reactions are a key diagnosis for the efficiency of iso and n-PrCN formation and may give insight into the observed abundances of both isomers. This information can then be useful for astrochemical modelling.\\

 The investigated reactions based on a selection of detected radicals in the ISM in gas phase are: \\ 

  \noindent CH$_3$CHCH$_3$+CN $\to$ iso$-$PrCN\\
  CH$_3$+CH$_3$CHCN  $\to$ iso$-$PrCN \\
  CH$_3$CH$_2$+CH$_2$CN $\to$ n$-$PrCN\\
  CH$_3$+CH$_2$CH$_2$CN $\to$  n$-$PrCN\\
   CH$_3$CH$_2$CH$_2$  +CN$\to$  n$-$PrCN\\

It is worth mentioning that the proposed radical-radical addition reactions we have selected were also considered as 
 main reactions included in the model employed by \citep{Belloche:14}.   
On the water ice-model, we also study the feasibility of the iso-PrCN and n-PrCN formation following the Eley-Rideal mechanism where gas phase species approach an already adsorbed radical. The choice of which of the two species is adsorbed first is based on the iso-PrCN and n-PrCN products binding energies on the ice mantle as will be explained in the main text.\\

 \noindent CH$_3$CHCH$_3$(gas)+CN$_{ads}$ $\to$ iso$-$PrCN$_{ads}$\\
  CH$_3$(gas)+CH$_3$CHCN$_{ads}$  $\to$ iso$-$PrCN$_{ads}$ \\
  CH$_3$CH$_2$(gas)+CH$_2$CN$_{ads}$ $\to$ n$-$PrCN$_{ads}$\\
  CH$_3$(gas)+CH$_2$CH$_2$CN$_{ads}$ $\to$  n$-$PrCN$_{ads}$\\
   CH$_3$CH$_2$CH$_2$(gas) +CN$_{ads}$  $\to$  n$-$PrCN$_{ads}$\\

In Section \ref{PES} we describe the adopted computational methods for electronic structure and rate constants calculations and in Section \ref{result} we discuss the results and their implications. Finally conclusions are given in Section \ref{conclusion}.

\section{Computational details}
\label{PES}

\begin{figure}
 \begin{center}
 \includegraphics[width=7.5cm] {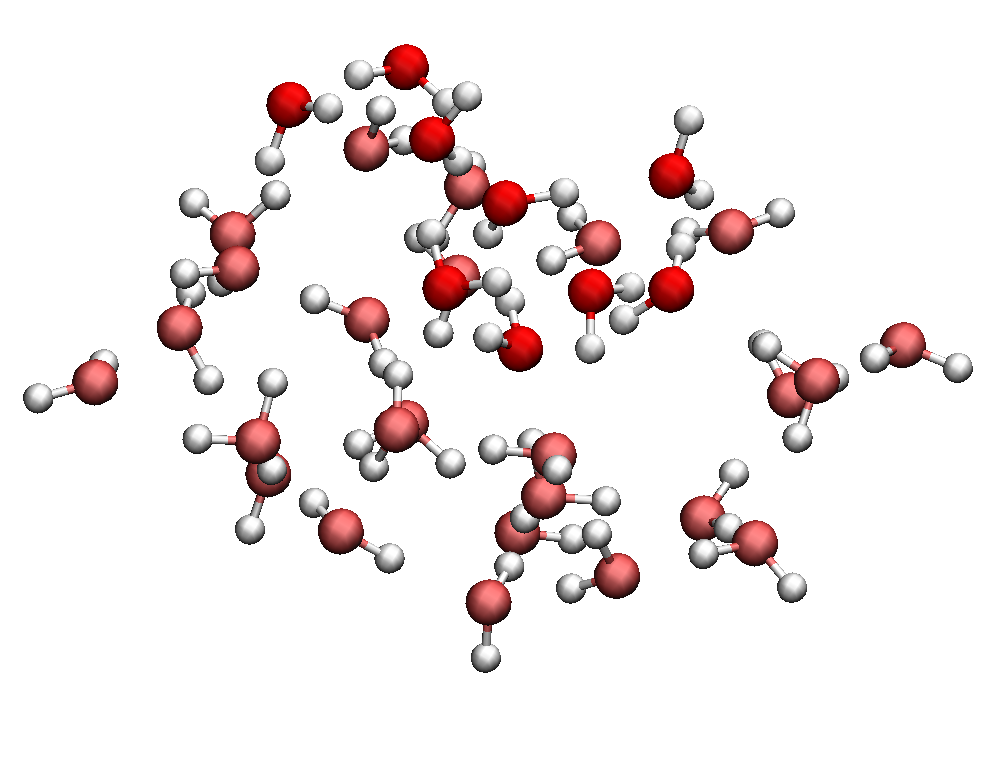}

  \caption{The 34 amorphous water-ice model built in this work fully optimised using M062X/6-311++g(d,p).}
\label{Figure:1}
 \end{center}
 \end{figure}

In the present work we employ density functional theory (DFT) \citep{Hohenberg:1964} to investigate iso-PrCN and n-PrCN formation in gas phase and on an amorphous model water-ice cluster comprised of 34 water molecules see (Figure~\ref{Figure:1}). Furthermore, in gas phase we employ the second order M$\phi$ller Plesset perturbation theory (MP2) [see \citep{Kerkeni:19}] and single point energy calculations were performed using the explicitly correlated coupled cluster theory. Restricted and unrestricted DFT and MP2 methods were applied to the closed-shell and open-shell species, respectively.

All DFT and MP2 quantum chemical calculations have been performed with the Gaussian 16 software \citep{Gaussian2016}.

\subsection{Density Functional Theory (DFT) Calculations}
 
In our work all structures involved in the radical-radical association on the 34-water ice-model are fully optimized and the harmonic frequencies computed using DFT.  
Frequency calculations were performed in order to verify the minimum energy character of the involved structures.

Specifically, we have used the hybrid-meta GGA functional M062X \citep{Zhao:2008}. The exchange-correlation functional was employed in connection with the split valence triple $\zeta$ Pople basis set 6-311++G(d,p) \citep{Martin:1989} which includes polarisation and diffuse functions and is able to describe van der Waals interactions inside the ice model and molecular interactions with the fragments. Further, this basis set is shown to provide near experimental accuracy for many chemical reactions \citep{Singh:2016a,Singh:2016b}. 
 
Gaussian 16 automatically includes an ultrafine integration grid in the DFT calculations in order to improve the accuracy of the results. The grid greatly enhances the accuracy at reasonable additional cost. 

\subsection{Moller-Plesset (MP2) Calculations}

For gas phase species we employed second order M$\phi$ller Plesset perturbation theory (MP2) (see \citep{Kerkeni:19}) for geometry and frequency calculations too, hence reaction energies, and reaction paths can be improved with respect to those from DFT. We specifically employ the MP2/aug-cc-pVTZ (see \citep{Kerkeni:19}) model chemistry.\\

\subsection{Coupled Cluster Calculations}
Searching for a more accurate description of the energetics, for all involved species in gas phase processes that were optimized with MP2/aug-cc-pVTZ level of theory, single point energy calculations were performed using the explicitly correlated coupled cluster theory \citep{knowles:93, knowles:00,knizia:09}, CCSD(T)-F12 implemented in \textsc{molpro} \citep{MOLPRO}  using the corresponding default options. For this purpose, the aug-cc-pVTZ (AVTZ) basis sets of atomic orbitals \citep{dunning:89,woon:93} were employed in connection with the corresponding basis sets for the density fitting and the resolution of the identity. Relative energies were refined using CCSD(T)-F12/AVTZ.
    
   \subsection{Rate Constants Calculations}
   All studied reactions either in gas phase or on the ice-model are barrier-less processes, therefore they do not pocess a well-defined transition state, hence the potential energy curves were fitted to a Varshni function \citep{Varshni:1957}:
   
   \begin{equation} D_e [1-\frac{R_e}{R}\{\exp[-\beta(R^2-R_e^2)]\}]^2\end{equation} 
    whose arguments are the three parameters D$_e$ (cm$^{-1}$), $\beta$ ($\AA^{-2}$) and R$_e$ (\AA). The long-range part of the Varshni potential may give a more faithful representation of the bond stretching curve than the Morse curve 
\citep{klippenstein:1988,Varshni:1957}.

The thermal Low-Pressure-Limit (LPL) rate constants for the gas phase reaction mechanisms have been computed with VRC-TST theory using the Variflex code \citep{VARIFLEX} in the 10-300 K temperature range and at pressures of 10$^{-7}$ torr. They are obtained by averaging over all J values by considering a thermal angular distribution of the complex \citep{Forst:1999}:

\begin{equation}
k_{\infty} (J,T) = \frac{\int_{0}^{\infty} k(E,J) N(E,J) \exp(-E/kT) dE}{ \int_0^{\infty} N(E,J) exp(-E/kT)dE}
\end{equation}

The reaction energies, rotational constants, and vibrational frequencies were firstly computed with the MP2/aug-cc-pVTZ method and then used in this rate constant calculation. The number of states of the "transition state" was energy E and angular momentum J (E/J ) resolved level \citep{Klippenstein:1994}. It is composed of two contributions: one is from the conserved modes which have little changes from reactants to products and are treated as harmonic vibrators employing the frequencies and geometries of the isolated fragments. The other one is from the transitional modes that shape-up during the reaction such as rotations and bending. The flexibility of the inter- fragment modes was treated for with the VRC using the Monte Carlo integration of a classical phase space representation.  This approach is used in many works to evaluate rate constants between complex species, see for example \citep{Zhou:2010,Pham:2020}.

\section{Results}
\label{result}
\subsection {Electronic structure calculations}
\subsubsection {Gas Phase Reactions}
\begin{figure*}  
 \begin{center}
  \par{
     \includegraphics[width=6cm]{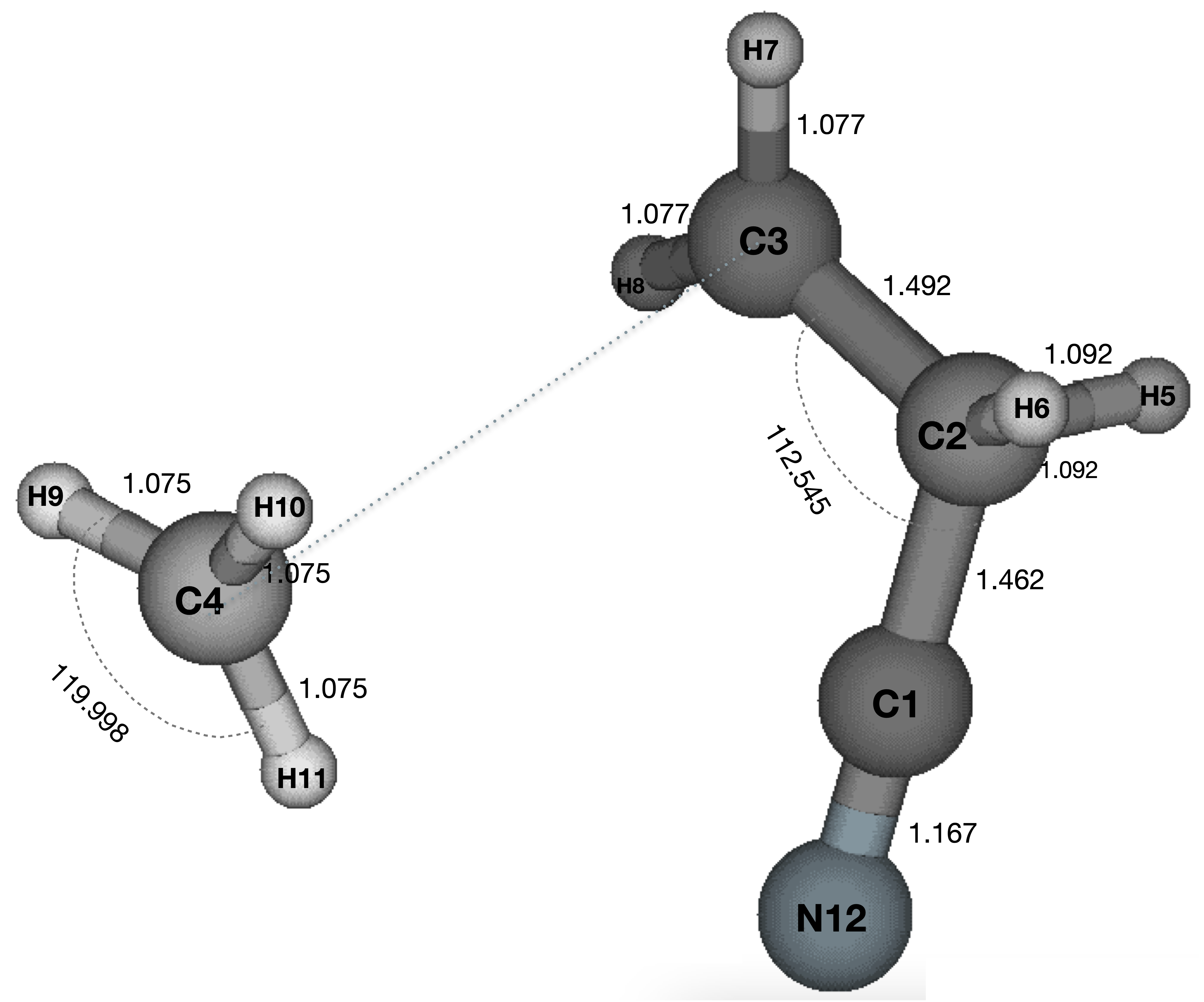}
    \includegraphics[width=7cm]{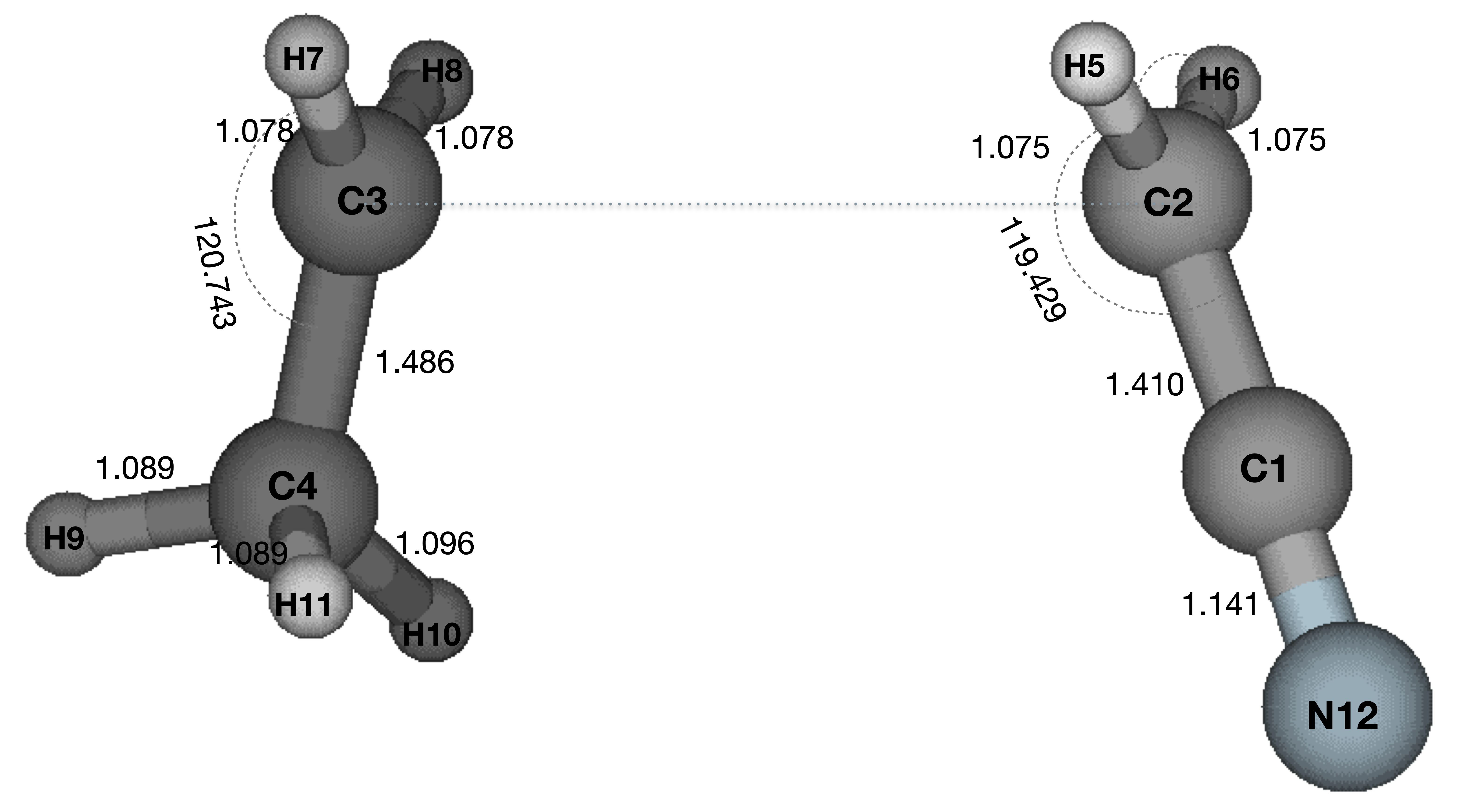}
 \includegraphics[width=5cm]{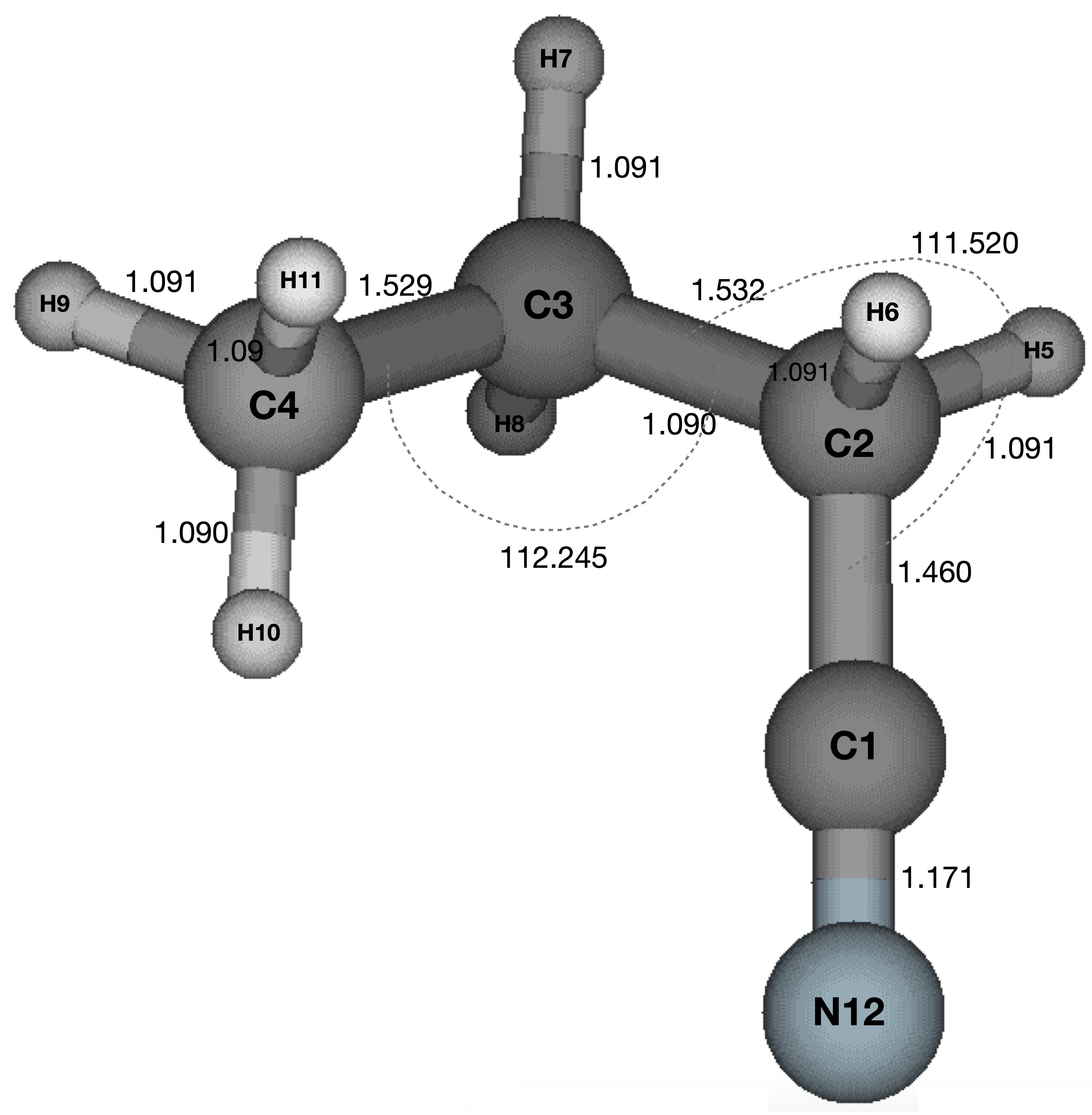}
      \includegraphics[width=6cm]{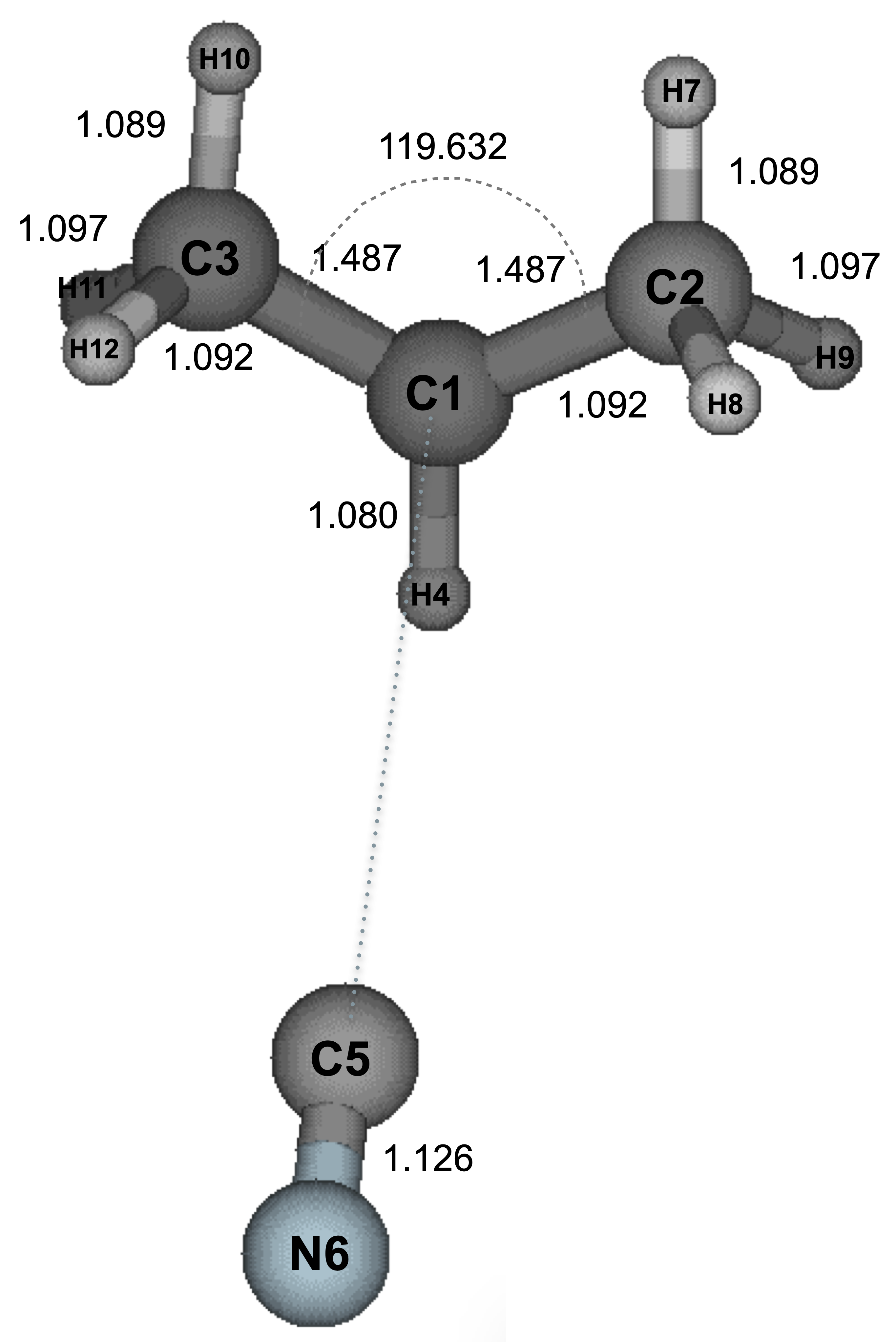}
          \includegraphics[width=5cm]{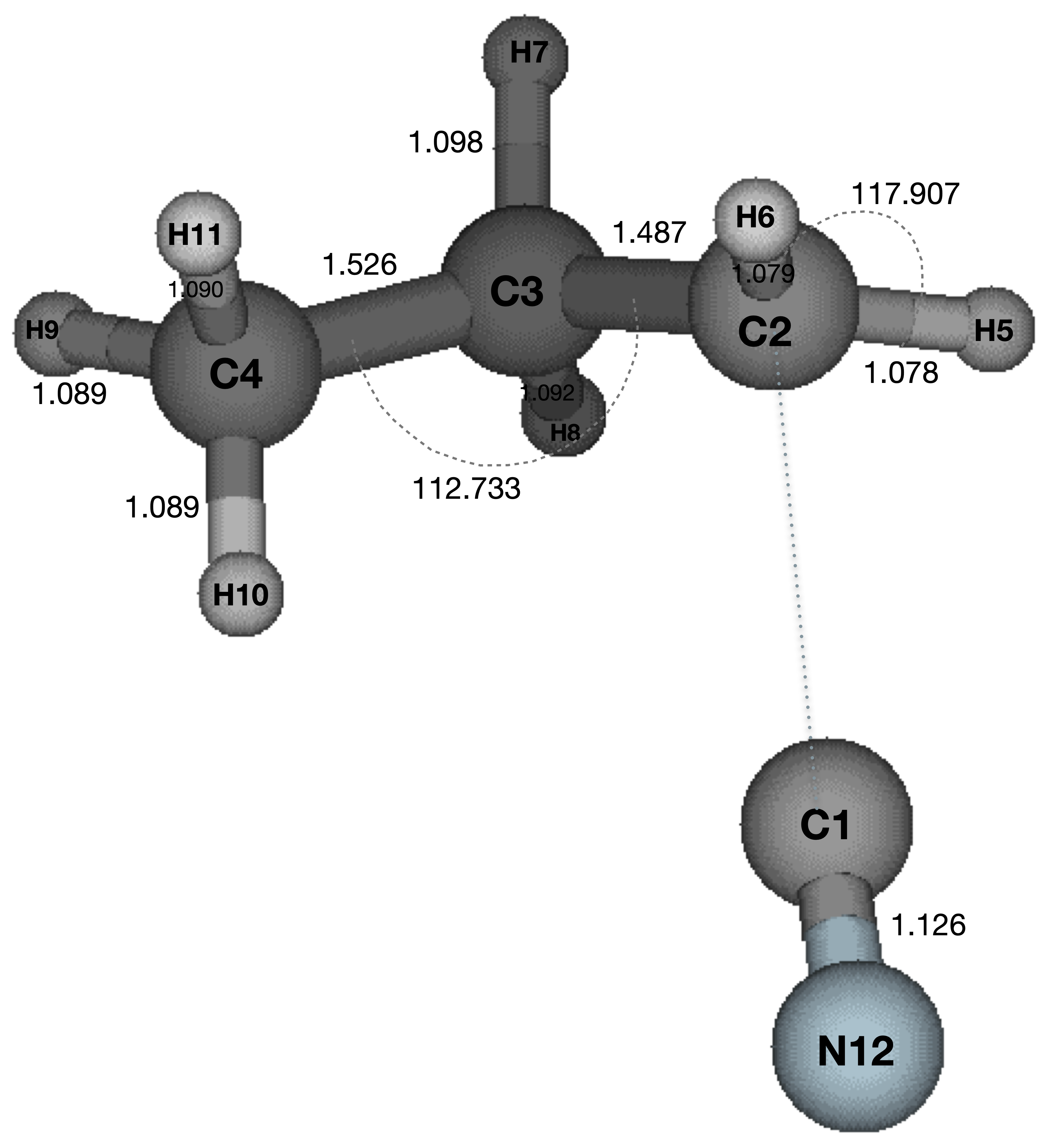}  
   \includegraphics[width=7cm]{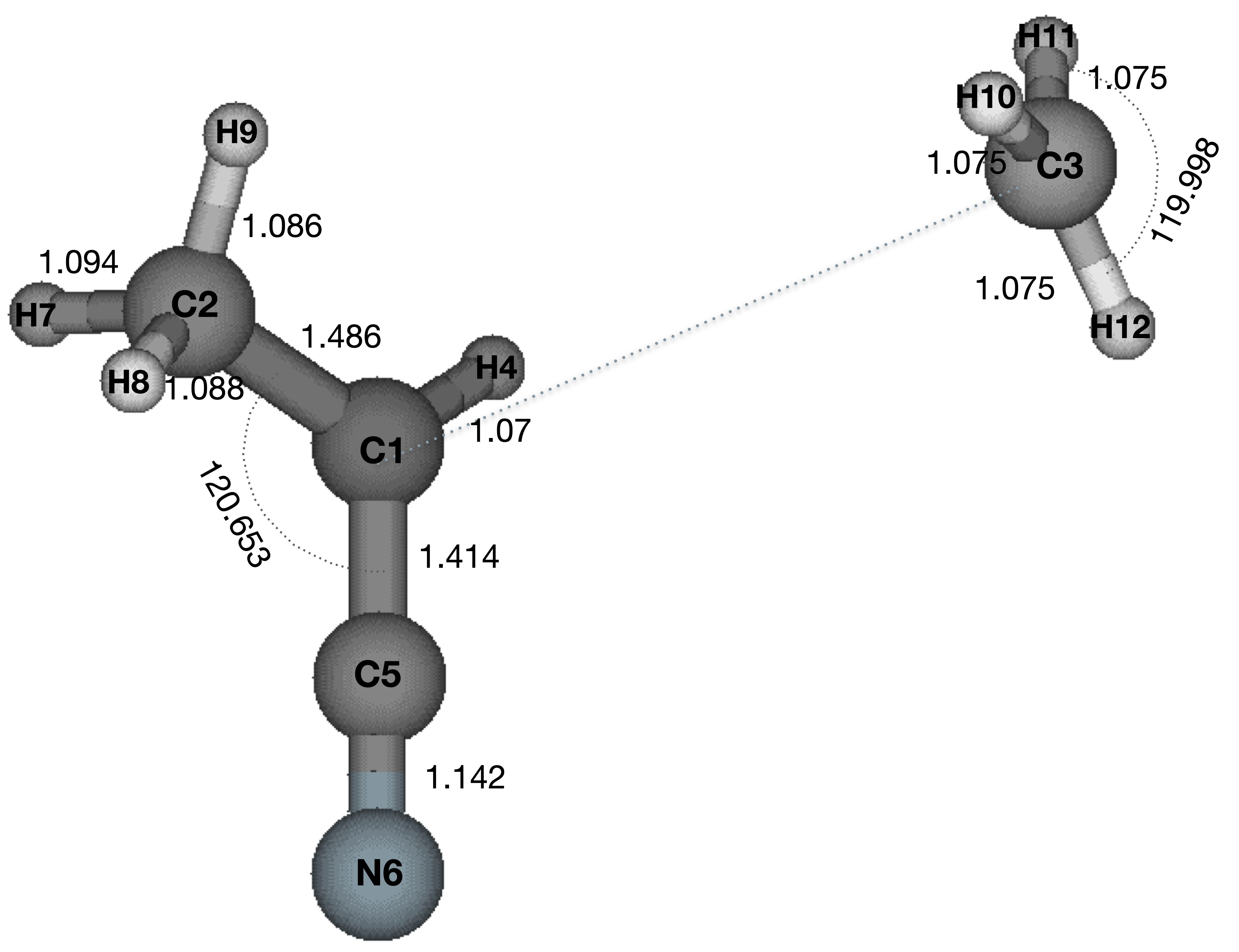}
 \includegraphics[width=6cm]{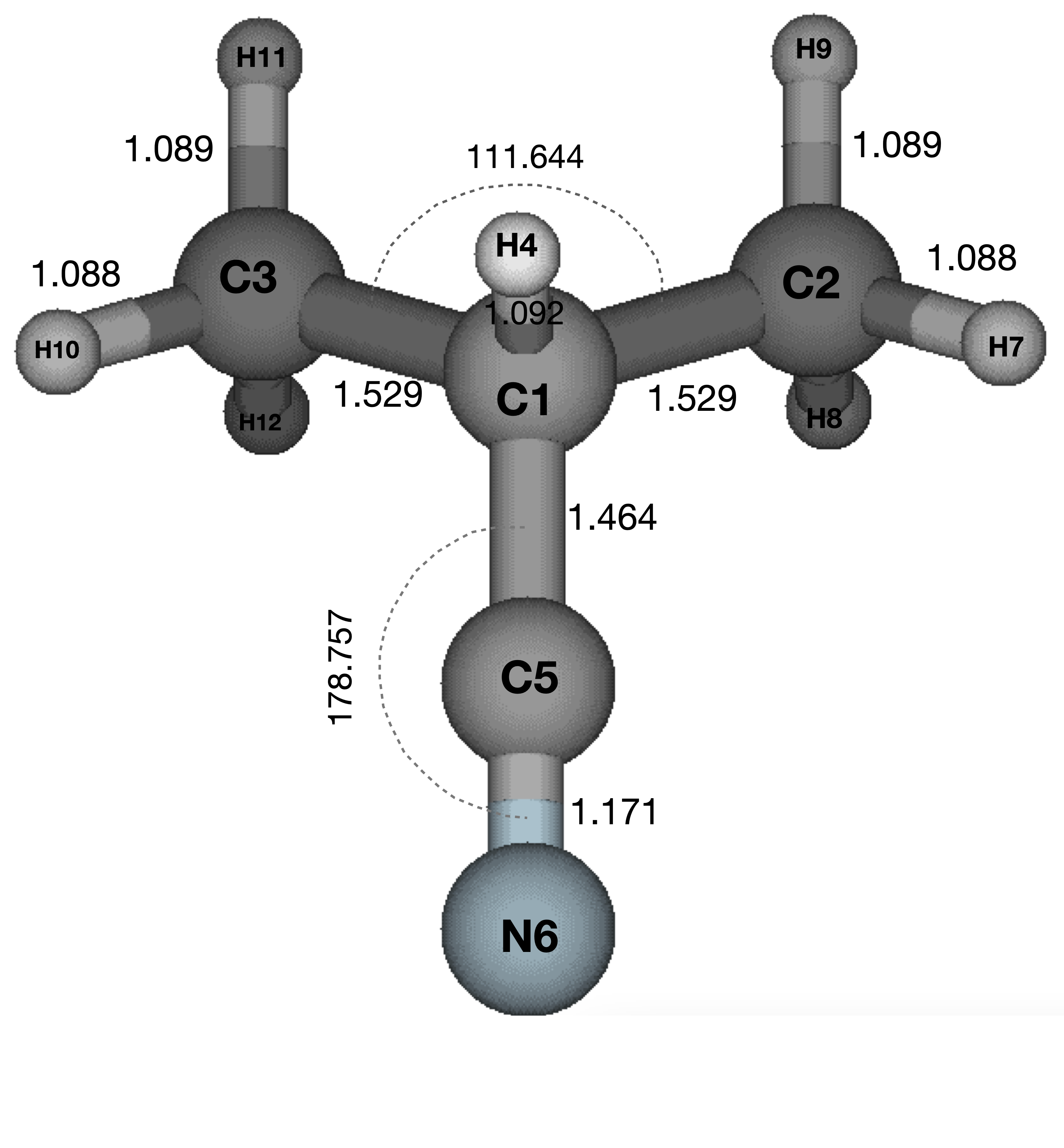}

\par}
 \caption{Geometries of the fully optimised different species, products (iso-PrCN, n-PrCN) optimized at MP2/aug-cc-pVTZ. The atoms that form the bond are connected with a dashed line.}\label{Figure:2}
 \end{center}
 \end{figure*}

 From the electronic structure calculations, we computed radicals, reactants, and products energies and frequencies.
All five reactions are found to be exothermic either with M062X/6-311++g(d,p) or MP2/aug-cc-pVTZ. For all fragments and products, we perform global optimisations of the coordinates with MP2/aug-cc-pVTZ level of theory which are displayed in Figure~\ref{Figure:2}. All reaction mechanisms are barrier-less and no transition states were encountered. The reaction paths are computed by specifying the required coordinate to be scanned with the Opt=Modredundant keyword and the remainder of the atoms were fully relaxed, allowing single point energy evaluations to yield a relaxed potential energy surface (PES) along the forming-breaking coordinate. The resulting five potential energy values and the fit with the Varshni equation (whose parameters are given in the appendix Table~\ref{table:4}) are plotted in Figure~\ref{Figure:3}.
 

To assess the adequacy of the MP2 methodology in predicting reaction energies, we further refine them by computing CCSD(T)-F12 single point energy calculations. Comparison between M062X, MP2 and CCSD(T)-F12 reaction energy and enthalpy
of reaction (including the difference of zero-point vibrational
energy $\Delta$(ZPE) between the product and the reactants can be seen in Table~\ref{table:1}. One can notice that in some cases MP2 overestimates the reaction energies with respect to M062X by as much as $\sim$ 17kcal/mol as expected.
Here, the MP2 reaction energies values vary monotonically with CCSD(T)-F12 one as already observed in the previous work employing CCSD(T) (see Table 4 of \citep{Kerkeni:19}). The CCSD(T)-F12 reaction energies should be of better accuracy and surprisingly the calculated values approach the M062X ones. At all levels of theory reactions with the CN radicals are more exothermic than the other CH$_3$ and C$_2$H$_5$+CH$_2$CN counterparts, with a maximum difference ranging from $\sim$ 40 kcal/mol for M062X and CCSD(T)-F12 and $\sim$ 52 kcal/mol for MP2.
The energy difference between n-PrCN (gauche) and n-PrCN (anti) is 0.37 and 0.06 kcal/mol at the MP2 and CCSD(T)-F12 levels respectively, in the remainder of the paper we only consider iso-PrCN and n-PrCN(gauche) which are the two most stable isomers for adsorption on the ice-model and for velocities calculations. \\

\subsubsection {Reactions on the ice model}

The ice model comprised of 34 water molecules was initially globally optimised with the M062X/6-311++g(d,p) method and the final geometry is given in Figure~\ref{Figure:1}. This combination of method and basis set gives satisfactory accuracy for the thermochemistry, and non-covalent interactions \citep{Peverati:2011}. Thus, it can be used to study our model system and describes accurately the interactions that play an important role in the stability of the different fragments involved in the reaction processes leading to the two products iso-PrCN and n-PrCN \citep{Chai:2008}.

 \subsubsection {Adsorption energies}

\begin{figure*}
 \begin{center}
 \par{
\includegraphics[width=4.5cm] {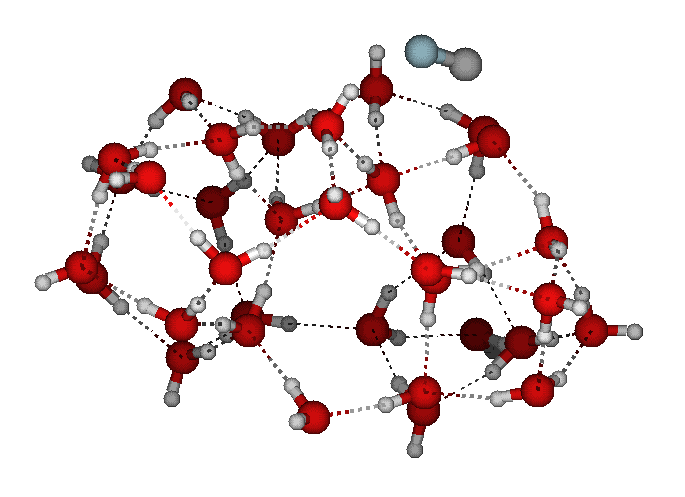}
\includegraphics[width=4.5cm] {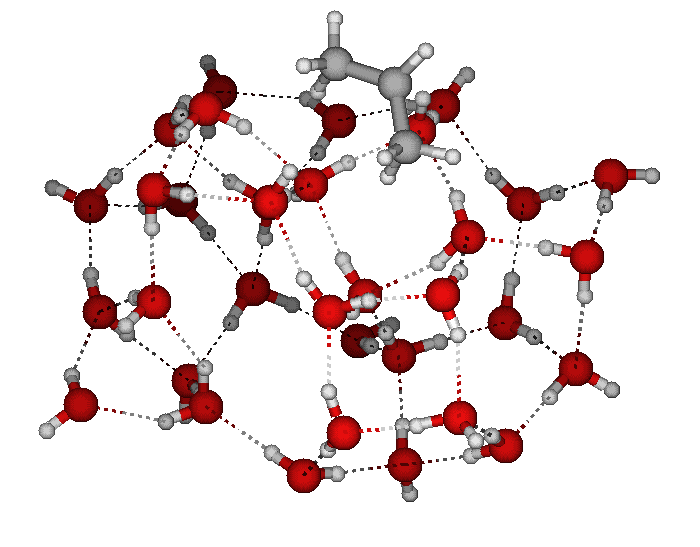}
\includegraphics[width=4.5cm] {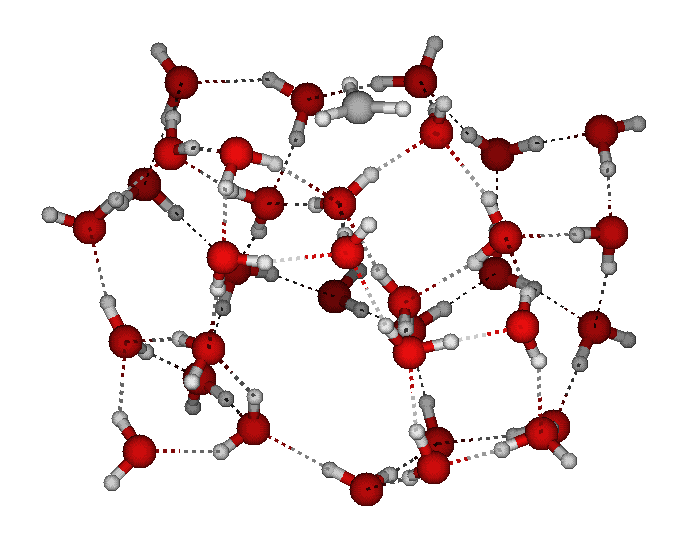}
\includegraphics[width=4.5cm] {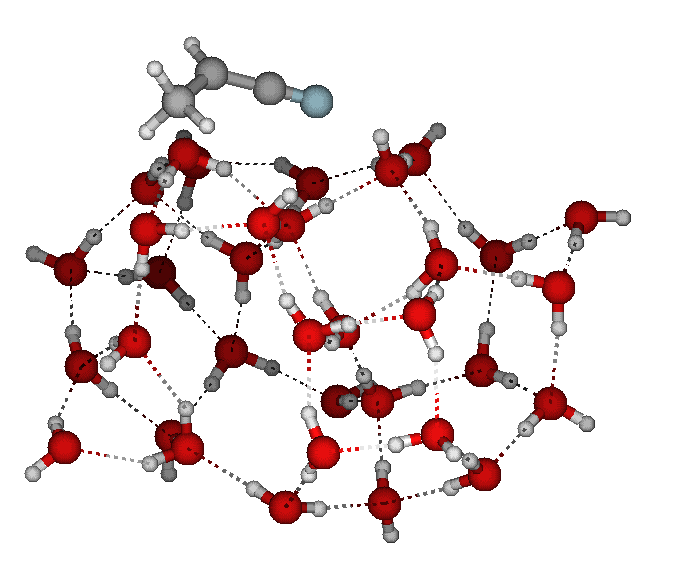}
\includegraphics[width=4.5cm] {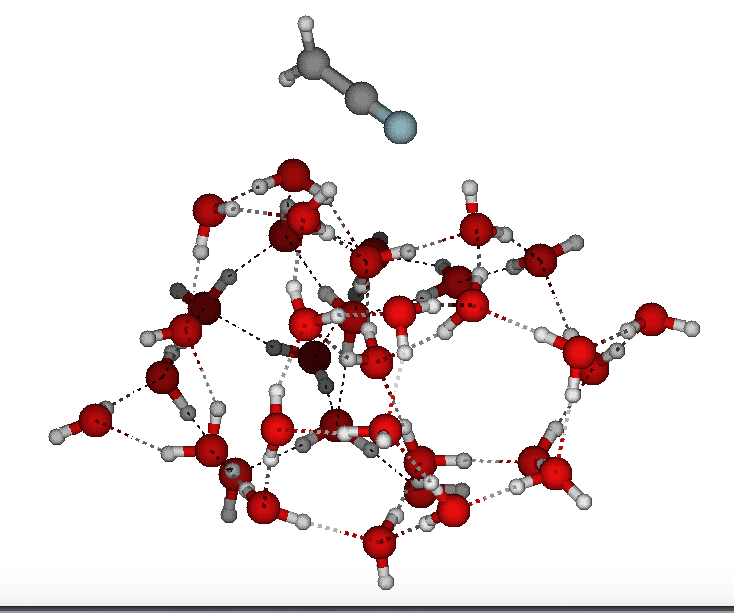}
\includegraphics[width=4.5cm] {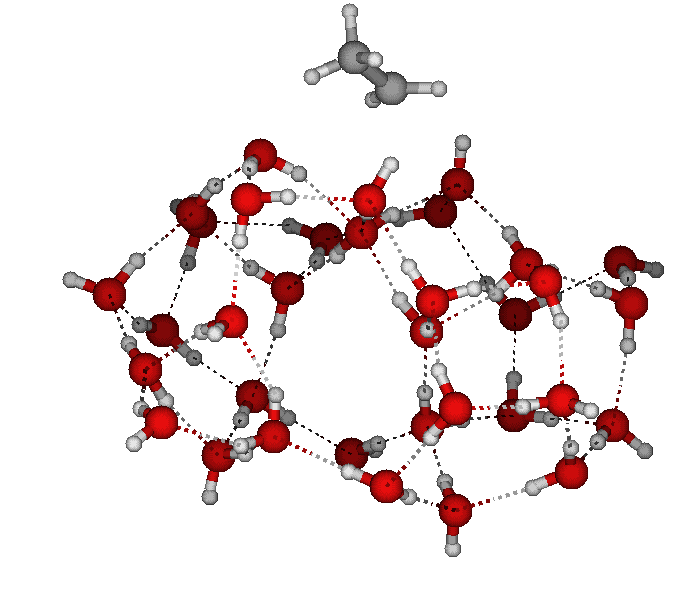}
\includegraphics[width=4.5cm] {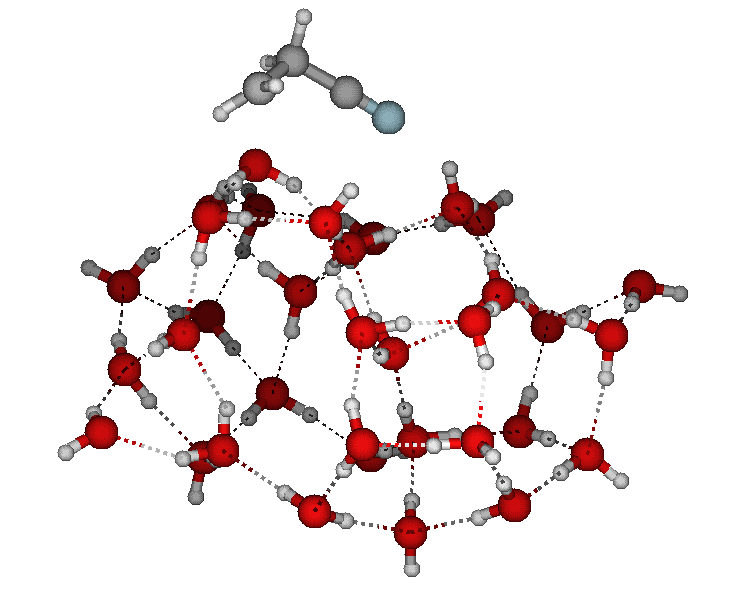}
\includegraphics[width=4.5cm] {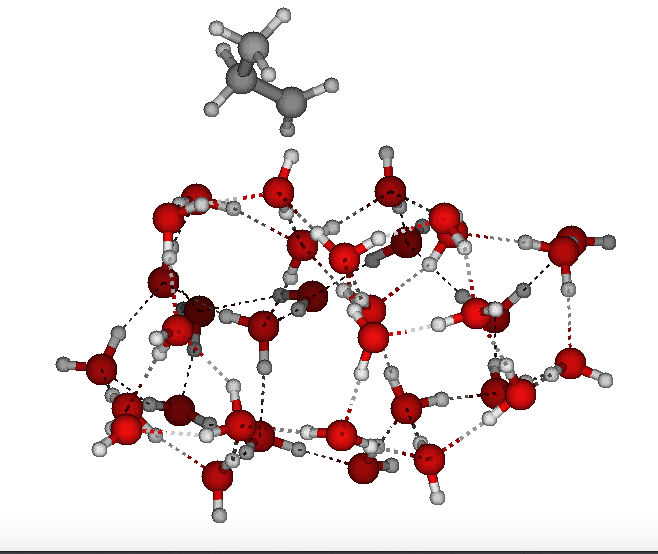}
\includegraphics[width=4.5cm] {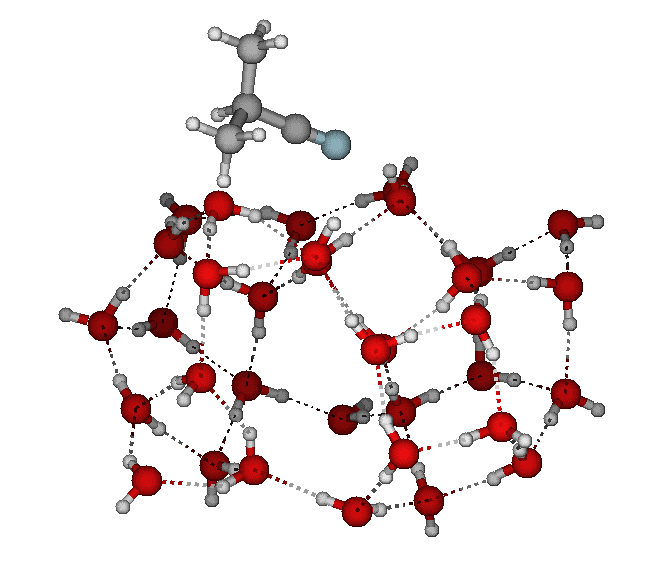}
\includegraphics[width=4.5cm] {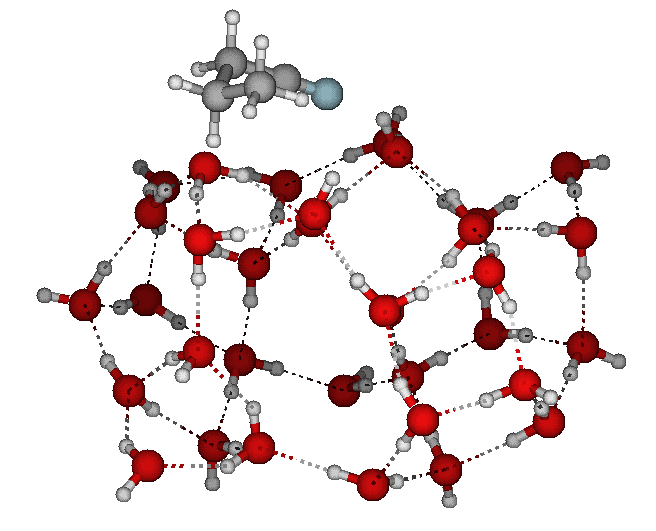}
\par}
  \caption{Optimized final geometry for the different adsorbed species: CN, CH$_3$CHCH$_3$, CH$_3$, CH$_3$CHCN, CH$_2$CN,  CH$_3$CH$_2$, CH$_2$CH$_2$CN, CH$_3$CH$_2$CH$_2$, iso-PrCN, and n-PrCN on the ice model using M062X/6-311++g(d,p).}
\label{Figure:4}
 \end{center}
 \end{figure*}

 The adsorption energy is computed as follows:  \begin{equation} E_{ads}=E(cluster)+E(adsorbate)-E(total)\end{equation}
  where E(cluster) is the total energy of the water-ice cluster model, E(adsorbate) is the energy of the gaseous species, and E(total) the energy of the adsorbed species upon approach on the ice. 
  In these calculations, we have allowed for the nine nearest H$_2$O molecules to the adsorbate to relax in geometry optimisations and have computed the frequencies to be sure we have a minimum potential energy. 
  In Figure~\ref{Figure:4} are displayed the optimised structures of the adsorbed radicals and also iso-PrCN and n-PrCN on the 34-waters ice model using M062X/6-311++g(d,p). The adsorption energies are shown and compared in Table~\ref{table:2}. 
 Several configurations of iso-PrCN and n-PrCN have been investigated in the quantum chemistry calculations, such as the CN group or the C$_3$H$_7$ moiety is facing the ice model. The most bound geometry is the one depicted in Figure~\ref{Figure:4} which corresponds to a CN group pointing to the ice-model in the case of the two products. Therefore, dissociation of the iso-PrCN and n-PrCN products was considered starting from their equilibrium most stable adsorbed position. On the other side, investigation of the adsorption energies of each fragment involved in the five radical-radical reactions supports this choice, as confirmed by the analysis of their separate binding energies. From Table~\ref{table:2} it is inferred that:\\
 \noindent  $-$CH$_3$CHCH$_3$ is as stable as CN within a fraction of a kcal/mol.\\
  \noindent $-$CH$_3$CHCN is more stable than  CH$_3$ by 7 kcal/mol.\\
  \noindent $-$CN is more stable than  CH$_3$CH$_2$CH$_2$ by 1.1 kcal/mol.\\
  \noindent $-$CH$_2$CH$_2$CN is more stable than  CH$_3$ by 1.6 kcal/mol.\\
  \noindent $-$CH$_2$CN is more stable than  CH$_3$CH$_2$ by 4 kcal/mol.\\

All these binding energies confirm our choice for the incoming gaseous radical and the accreted one in the association reactions involving the ice-model. We provide in Supplementary Material the cartesian coordinates for all 5 adsorbed fragments and also those relevant to iso-PrCN and n-PrCN adsorbed on the 34 amorphous water-ice model computed with M062X/6-311++g(d,p) model chemistry. 

\subsubsection {iso-PrCN formation}

The formation route of iso-PrCN may involve the addition of the CN radical to CH$_3$CHCH$_3$ or the methyl addition to CH$_3$CHCN moiety. Investigations of the adsorption energies shown above for the four involved species demonstrate that CN is as stable as CH$_3$CHCH$_3$ upon adsorption on the ice mantle and that CH$_3$CHCN is more stable than CH$_3$. Furthermore, the formation of iso-PrCN occurs without a barrier.

\subsubsection {n-PrCN formation}
The chemical mechanisms leading to the formation route of n-PrCN may involve the addition of the CH$_3$CH$_2$ radical to CH$_2$CN, the methyl addition to CH$_2$CH$_2$CN moiety, or the CN radical addition to CH$_3$CH$_2$CH$_2$. Investigations of the adsorption energies of the four mentioned species as detailed above and also from Table~\ref{table:2} confirm that CH$_2$CN is more stable than CH$_3$CH$_2$ upon approach to the ice mantle and that CH$_2$CH$_2$CN is as stable as CH$_3$, as is the case for CH$_3$CH$_2$CH$_2$ and CN. The formation of n-PrCN following the three reaction paths occurs without a barrier. 

\subsubsection{Reaction mechanisms} 
\label{reaction}
In the radical addition reactions on the ice model, where a first radical is adsorbed on ice and a second radical comes from the gas phase, we started the single point energies calculations from the geometry of the stable adsorbed product (iso-PrCN, or n-PrCN). In these calculations, all the ice-model water coordinates were kept frozen at their initially optimised geometry. All the geometries of the radicals that participate in the reaction are fully relaxed. In this case, the bonding coordinate between the gas phase and the adsorbed species atoms is varied in a range from 0.4-4 (\AA) to obtain the potential energy function. All reactions are exoergic and barrier-less and in Tables ~\ref{table:3} and  ~\ref{table:4} respectively, we collect the computed reaction energies, and the Varshni parameters. 
Due to the fact that the ice-model geometry is not symmetric with respect to the adsorbed radicals, and in particular the case of CH$_3$CH(CN), the addition of the two radicals CH$_3$ (left: L and right: R) has been explored, when they approach the same C atom from the gas phase. This has lead to nearly identical PESs, and hence only one is considered in this paper.

  \subsubsection {Reaction Rate Constants}
  
\begin{figure*}  
 \begin{center}
  \includegraphics[width=11.5cm] {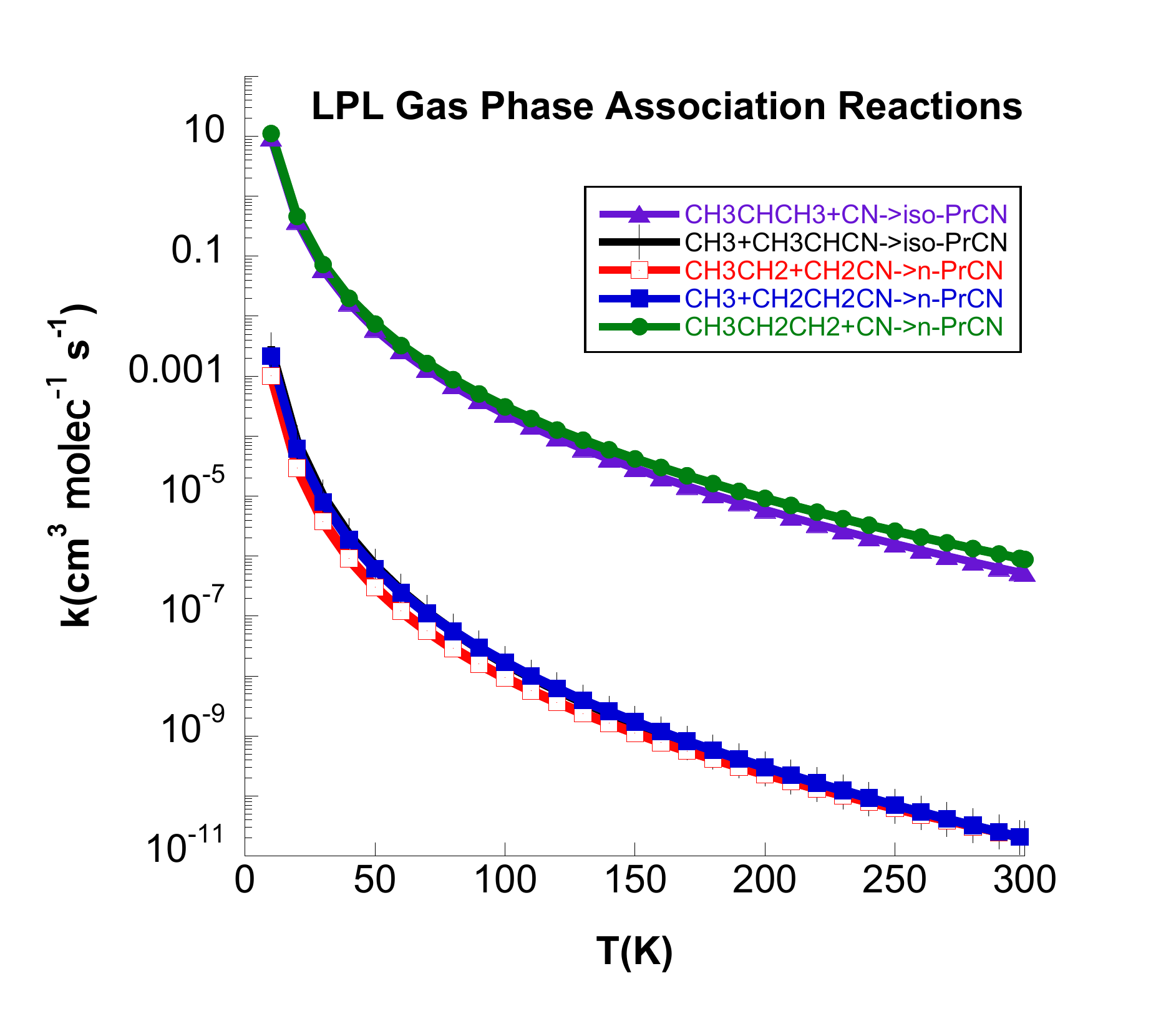}

  \caption{LPL Temperature dependent rate constants for the gas phase association reactions considered in this study.}

\label{Figure:5}
 \end{center}
 \end{figure*}

One major goal of this study is to compute rate constants for association reactions leading to iso-PrCN and n-PrCN for the 5 gas phase reactions detailed above. The rate constants in the LPL were computed in the temperature range 10-300K relevant to cold and hot core ISM regions. The total angular momentum J covered the range from 3 to 351 in steps of 12 for all the E, J-resolved calculations. In order to achieve convergence in the integration over the energy range, an energy step size of 50 cm$^{-1}$ was used.\\
The anisotropy of the potential curves was taken into account by specifying the angles corresponding to the equilibrium structures of the two incoming radicals, the Lennard-Jones pairwise atom-atom interactions are taken into account.\\
The moments of inertia and vibrational frequencies of iso-PrCN and n-PrCN that are used in the rate constant calculations are presented in Table~\ref{table:5}.\\
A Varshni equation was fitted to all the relaxed scan data of the potential energy values calculated above to form the two products iso-PrCN or n-PrCN. The rate constants values are tabulated in Table~\ref{table:6} and plotted in Figure~\ref{Figure:5}. The half-lives $\tau$ also given in the same table show that the largest rate reactions exhibit shorter $\tau$ values.
The reactions display a negative temperature dependence which is usual for barrier-less recombination reactions \citep{Slagie:1988,Villa:1997}. The more exoergic reactions involving CN addition have the largest rate constants. These LPL rate coefficients are characteristic of the long-lived reaction complexes at lower temperatures as reported by \citep{Wu:2018} in the case of the collisional stabilization rate $k_s$ of their reaction complex. \\

The presently studied radical-radical association reactions are barrier-less, and the computed rate constants indicate that they are very efficient at the temperatures typical of the ISM. The low pressure limit of collisional association reveals that these reaction channels are open at all temperatures considered in this work and range between 10 and 10$^{-6}$ cm$^3$molec$^{-1}$sec$^{-1}$ for CH$_3$CHCH$_3$+CN $\to$ iso$-$PrCN and CH$_3$CH$_2$CH$_2$  +CN$\to$  n$-$PrCN, and between 10$^{-3}$ to 10$^{-11}$ cm$^3$molec$^{-1}$sec$^{-1}$ equally for   CH$_3$+CH$_3$CHCN  $\to$ iso$-$PrCN, CH$_3$+CH$_2$CH$_2$CN $\to$  n$-$PrCN, and CH$_3$CH$_2$+CH$_2$CN $\to$ n$-$PrCN.\\

\begin{table*}
\caption{Comparison between reaction energies (in kcal/mol) for the different reaction paths involved in iso-PrCN and n-PrCN formation in gas phase computed with M062X/6-311++g(d,p), MP2/aug-cc-pVTZ and compared to CCSD(T)-F12//MP2.}             
             
\label{table:1}      
\centering          
  \begin{tabular}{c|cc|cc|cc}
  \hline
     &  M062X   &&MP2&& CCSD(T)-F12& \\

  Species &  $\Delta$ E & $\Delta$ E+ $\Delta$ (ZPE)& $\Delta$ E & $\Delta$ E+ $\Delta$ (ZPE)&$\Delta$ E & $\Delta$ E+ $\Delta$ (ZPE)\\
  \hline

  CH$_3$CHCH$_3$+CN $\to$ iso$-$PrCN&-131.59&-125.60 &-148.4&-144.6&-127.3 &-122.9  \\
    \hline

  CH$_3$+CH$_3$CHCN  $\to$ iso$-$PrCN&-91.69&-83.39 &-101.5&-94.7&-89.6  &-82.8   \\
    \hline

  CH$_3$CH$_2$+CH$_2$CN $\to$ n$-$PrCN& -93.32& -85.31& -103.6&-97.1&-91.1&-84.5   \\
    \hline

  CH$_3$+CH$_2$CH$_2$CN $\to$  n$-$PrCN&-102.36& -93.47&   -102.5&-94.1&-98.4 &-90.0   \\
    \hline

   CN+CH$_3$CH$_2$CH$_2  $$\to$  n$-$PrCN&-134.51  & -128.31&  -150.7&-146.1&-129.6 &-124.9     \\
 
 \hline
\end{tabular}
\end{table*}

\begin{table*}
\caption{Adsorption energies (in kcal/mol) for the different species involved in iso-PrCN and n-PrCN formation on the 34-waters cluster computed with M062X/6-311++g(d,p) allvH$_2$O molecules in the ice are frozen.}             
\label{table:2}      
\centering          
  \begin{tabular}{c|c cc}
  \hline
 
  Species &   E$_{ads}$ & E$_{ads}$+ $\Delta$ (ZPE) \\
  \hline
  
CN& 7.8    &6.9  \\
 \hline
CH$_3$CHCH$_3$ &  8.1&6.8   \\
  \hline
CH$_3$&  4.5 &	3.3 \\
  \hline
  CH$_3$CHCN &11.7  &	10.9  \\
  \hline
CH$_3$CH$_2$ &6.2 &5.3	  \\
  \hline
CH$_2$CN  &  10.3 &9.5	  \\
  \hline
 CH$_2$CH$_2$CN  & 2.9&2.4  \\
  \hline
 CH$_3$CH$_2$CH$_2$& 6.6  &	5.9\\
   \hline
  iso$-$PrCN &  12.1 &	11.5\\
    \hline
  n$-$PrCN & 12.9 &12.3	 \\
 \hline
\end{tabular}
\end{table*}

\begin{table*}
\caption{Reaction energies (in kcal/mol) for the different reaction paths involved in iso-PrCN and n-PrCN formation on the 34-waters cluster computed with M062X/6-311++g(d,p), the whole set of the water molecules are frozen.}             
\label{table:3}      
\centering          
  \begin{tabular}{cccc }
  \hline
 
  Species &   $\Delta$ E & $\Delta$ E+ $\Delta$ (ZPE)\\
  \hline
  
  CH$_3$CHCH$_3$(gas)+CN$_{ads}$ $\to$ iso$-$PrCN& -131.8&-126.2\\
    \hline

  CH$_3$(gas)+CH$_3$CHCN$_{ads}$  $\to$ iso$-$PrCN& -88.2&-80.1 \\
    \hline

  CH$_3$CH$_2$+CH$_2$CN$_{ads}$ $\to$ n$-$PrCN& -92.1&-84.3\\
    \hline

  CH$_3$(gas)+CH$_2$CH$_2$CN$_{ads}$ $\to$  n$-$PrCN& -97.5&-88.8\\
    \hline

   CN$_{ads}$+CH$_3$CH$_2$CH$_2(gas)$ $\to$  n$-$PrCN& -135.8&-129.9\\
 
 \hline

\end{tabular}
\end{table*}

      \begin{figure*}  
 \begin{center}
  \par{
\includegraphics[width=6.5cm] {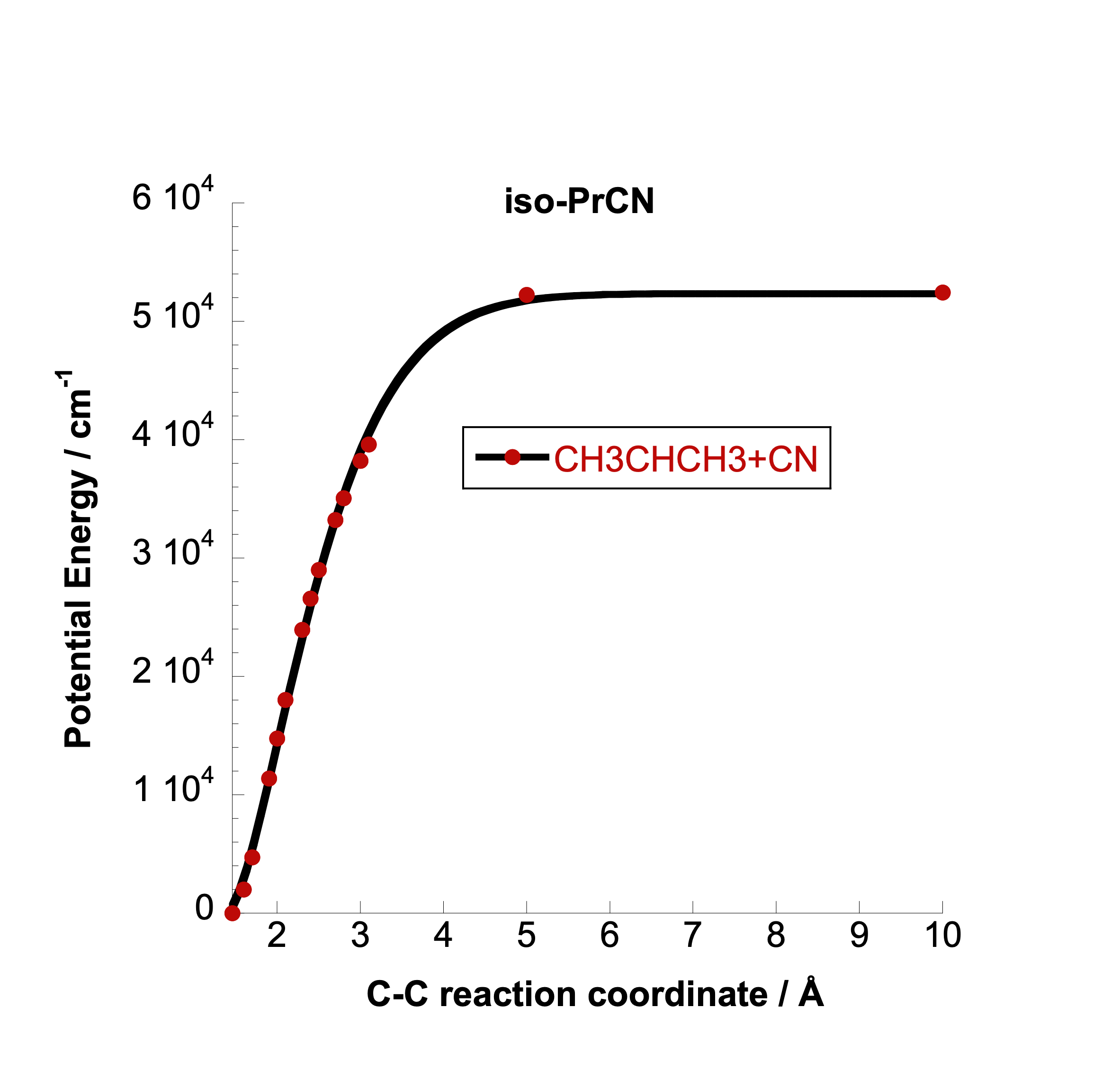}
\includegraphics[width=6.5cm] {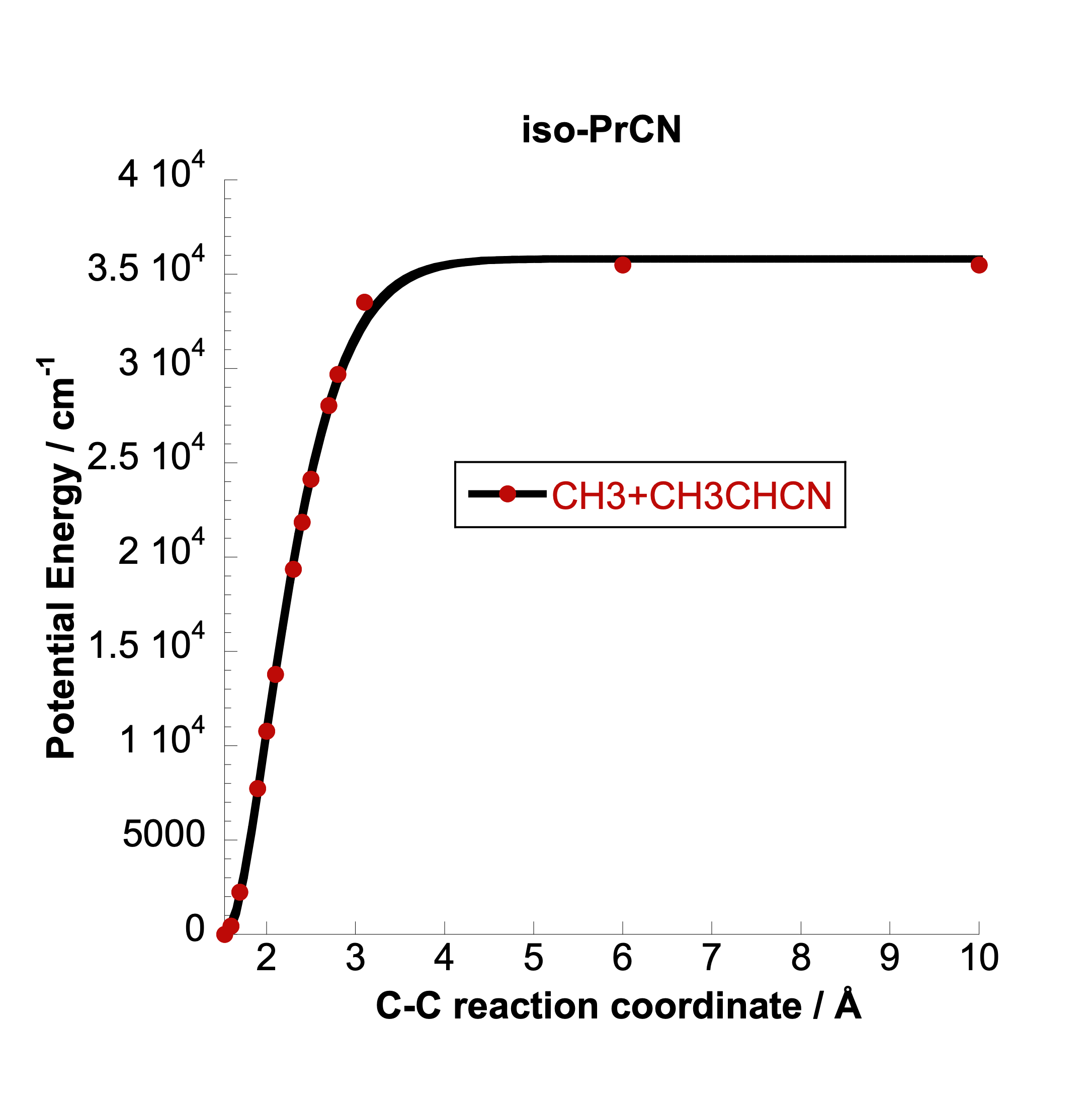}
\includegraphics[width=6.5cm] {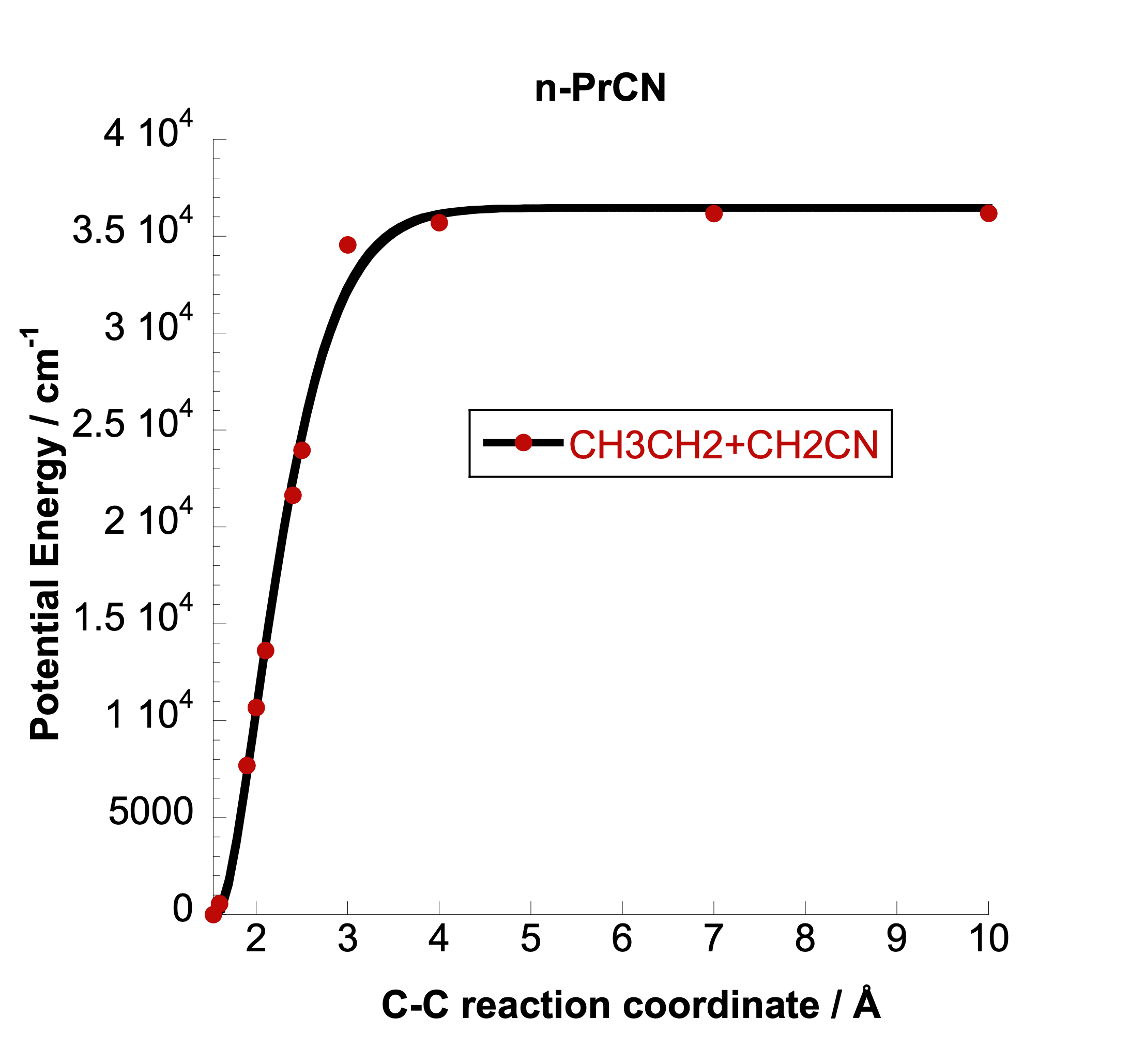}
\includegraphics[width=6.5cm] {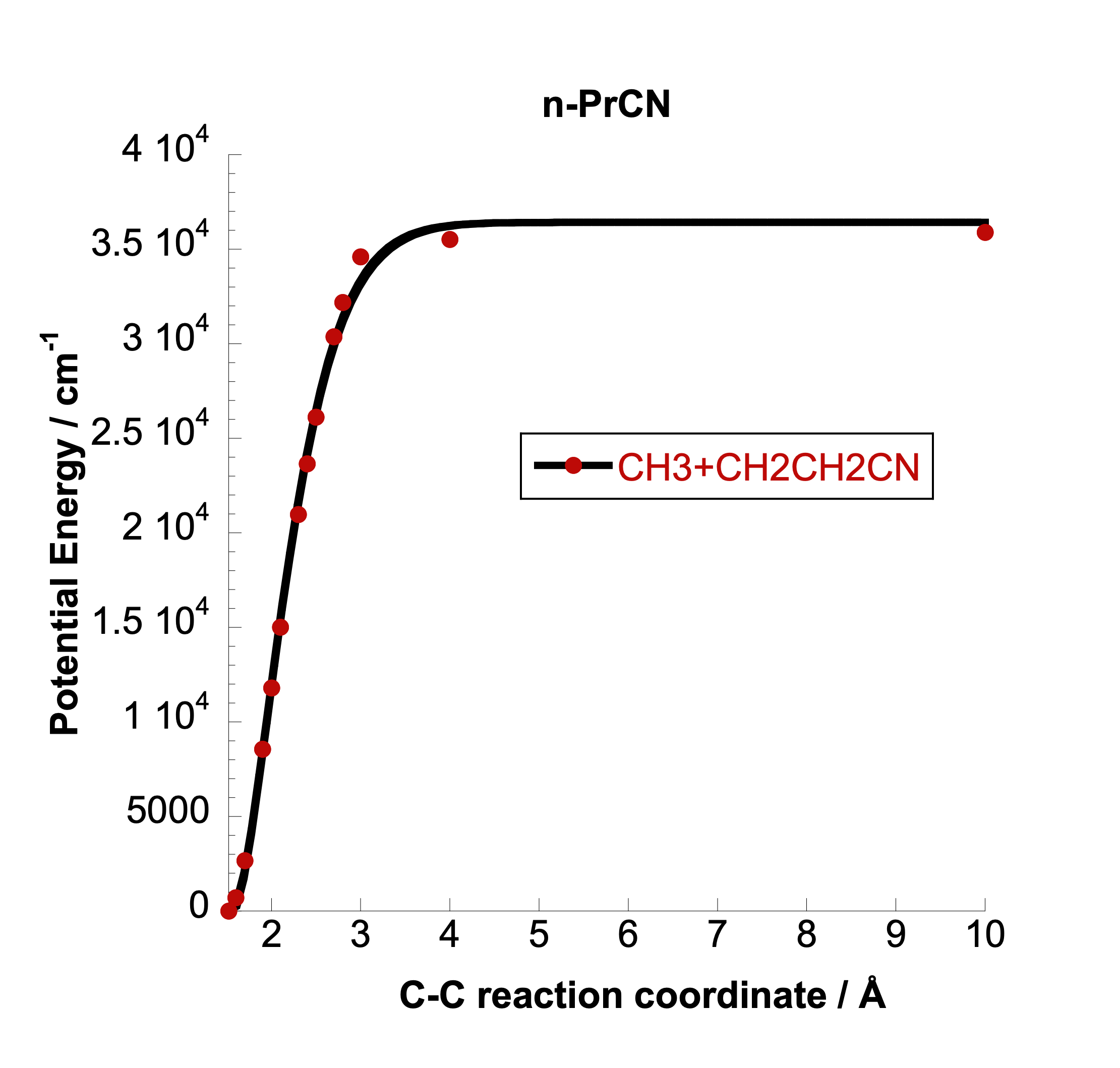}
\includegraphics[width=6.5cm] {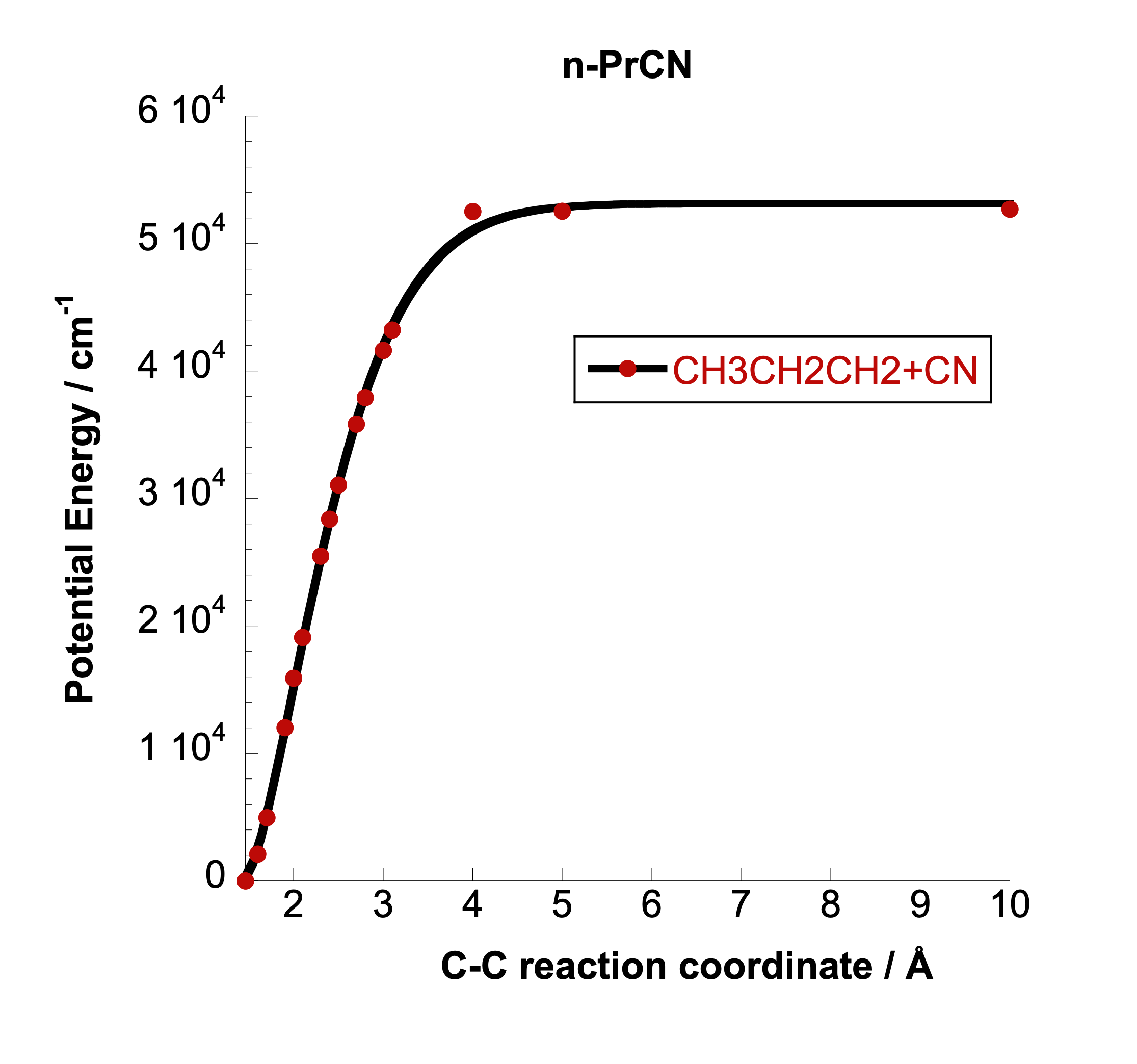}

\par}
  \caption{MP2/aug-cc-pVTZ Potential energy curves for radical-radical addition gas phase reactions.}
\label{Figure:3}
 \end{center}
 \end{figure*} 
 
\section{Summary and conclusion}

Gas phase and on the present ice-model, reaction mechanisms for the formation of propyl cyanide isomers have been investigated and we find that all reaction paths are barrier-less. All five channels are open for the temperature range considered.
DFT calculations have shown that the reaction energetics are similar in the gas phase and on the ice-model. Only in the gas phase MP2 and CCSD(T)-F12 energetics have been computed, and reaction paths calculated with MP2.
All association reactions are exothermic and the major channels for both isomers are found to be the CN radical addition to CH$_3$CHCH$_3$ and CH$_3$CH$_2$CH$_2$. The CH$_3$ addition and the C$_2$H$_5$ $+$ CH$_2$CN are less dominant mechanisms, as can be seen from the gas phase LPL rate coefficients. Furthermore, as can be seen from Table~\ref{table:3}, the association reactions on the ice involving addition of CN are the more exothermic and hence one would expect them to have larger LPL rate constants as was the case in the gas phase.
The detected abundance of iso-PrCN is $0.4$ times larger than that of its straight-chain structural isomer n-PrCN identified in the ALMA spectrum of the northern hot core of the star forming region Sgr B2(N). Rates have been calculated by \citep{Belloche:14} using the rate-equation approach as described by \citep{Garrod:13}.
The model employed shows that the greatest contribution to iso-PrCN production comes from the reaction of accreted gaseous CN
radicals as confirmed in our gas phase rate coefficient calculations of iso and normal isomers. While Belloche and coworkers find that the dominant formation mechanism for n-PrCN is the addition of C$_2$H$_5$ and CH$_2$CN. Later, \citep{Garrod:17} and \citep{Garrod:2022} modified the chemical network and have a different conclusion.
However, the lack of well-validated kinetic data for the formation of PrCN isomers hampers our understanding of the iso/n abundance in the ISM. 
The presently computed reaction rates may be useful in elucidating interstellar chemical networks.

\label{conclusion}

\section*{Acknowledgments}
BK is thankful to the COST Actions CM1401 "Our Astrochemical History" and to the Tunisian Ministry of Higher Education and Scientific Research for a visiting professorship to the CSIC. The authors acknowledge the Ministerio de Ciencia, Innovación y Universidades of Spain through the grant PID2020-112887GB-I00. MLS is thankful to the CTI (CSIC) and CESGA and to the “Red Española de Computación” for the grants AECT-2020-2-0008 and RES-AECT-2020-3-0011 for computing facilities. This work was also granted
access to the HPC resources of MesoPSL financed by the ”Région
\^Ile de France” and the project Equip@Meso (reference ANR-10-
EQPX-29-01) of the ”programme Investissements d’Avenir” supervised
by the ”Agence Nationale pour la Recherche”

\section*{Data Availability}

The following files are available free of charge.\\

COORD: ASCII files for all XYZ coordinates pertaining to the five considered species adsorbed on ice and optimised at the M062X/6-311++g(d,p) model chemistry. The MP2/aug-cc-pVTZ optimised gas phase radicals and products are also available.\\
  \noindent

\bibliographystyle{mnras}

\bibliography{PrCN-26-01-2023}
\appendix

\section{Some extra material}

\begin{table*}
\caption{Varshni parameters (D$_e$/cm$^{-1}$, $\beta$ / \AA$^{-2}$, and  R$_e$ / \AA) for the gas phase potential energy curves used in the rate constant calculations and to fit the data in Figure~\ref{Figure:3}. Those relevant to the radical-radical reactions on the ice mantle are also given.}             
\label{table:4}      
\centering          
  \begin{tabular}{cccc}
  \hline
  Gas phase reactions&    &  &   \\

 &   D$_{e}$& $\beta$&R$_{e}$   \\
  \hline
  
 CH$_3$CHCH$_3$+CN $\to$ iso$-$PrCN &  52245.1 & 0.199& 1.47  \\
 \hline
CH$_3$+CH$_3$CHCN  $\to$ iso$-$PrCN &  38725.8&  0.278& 1.53   \\
  \hline
CH$_3$CH$_2$+CH$_2$CN $\to$ n$-$PrCN&  36562.4 & 0.312 &1.53\\
  \hline
  CH$_3$+CH$_2$CH$_2$CN $\to$  n$-$PrCN &36634.5 & 0.349 &1.52 \\
  \hline
CN+CH$_3$CH$_2$CH$_2  $$\to$  n$-$PrCN &52638.3& 0.228& 1.46	  \\
  \hline
    On the ice reactions&    &  &   \\
  \hline
  CH$_3$CHCH$_3$+CN$_{ads}$ $\to$ iso$-$PrCN$_{ads}$ & 44949.9 & 0.215& 1.47  \\
 \hline
CH$_3$+CH$_3$CHCN$_{ads}$  $\to$ iso$-$PrCN$_{ads}$ &  31629.3&  0.409& 1.53   \\
  \hline
CH$_3$CH$_2$+CH$_2$CN$_{ads}$ $\to$ n$-$PrCN$_{ads}$&  32107.9  & 0.353 &1.53\\
  \hline
  CH$_3$+CH$_2$CH$_2$CN$_{ads}$ $\to$  n$-$PrCN$_{ads}$ & 35636.3  &0.355& 1.52 \\
  \hline
CN+CH$_3$CH$_2$CH$_2  $$_{ads}$$\to$  n$-$PrCN$_{ads}$ &47254.8& 0.189& 1.46	  \\
  \hline
\end{tabular}
\end{table*}

\begin{table*}
\caption{MP2/-aug-cc-pVZ Frequencies in cm$^{-1}$ and moments of inertia in cm$^{-1}$ .}             
\label{table:5}      
\centering          
  \begin{tabular}{ccccc}
  \hline
 
 &  & iso-PrCN&  &   \\
  \hline
     &179.6216& 214.4965& 228.4627&\\
& 286.9447& 351.6176& 539.5011&\\
& 557.0950& 783.9597& 943.5877&\\
& 950.3639& 988.1379&1134.4931&\\
&1149.4001&1205.6629&1331.2556&\\
&1357.6499&1407.8779&1428.3473&\\
&1505.9920&1507.1776&1519.6178&\\
&1529.1723&2184.6292&3074.0267&\\
&3074.5600&3090.0413&3165.2559&\\
&3168.2643&3173.3356&3174.0310&\\
  \hline

&0.2675936 & 0.13229901& 0.097227782&\\
  \hline
 &  & n-PrCN&  &   \\
   \hline

  &110.6089& 173.3962& 266.6279&\\
& 351.5922& 368.2047& 552.9055&\\
& 774.1407& 863.2831& 888.7352&\\
& 939.6867&1079.1617&1110.6799&\\
&1135.0714&1265.1992&1295.8315&\\
&1364.5049&1379.4236&1426.4368&\\
&1484.1911&1505.8190&1519.9436&\\
&1524.2345&2192.3624&3068.4404&\\
&3084.1186&3086.3856&3131.0468&\\
&3142.3861&3158.3507&3165.1173&\\
  \hline

&0.33326701 &0.1107113686& 0.091285273&\\
\\
  \hline
\end{tabular}
\end{table*}

\begin{table*}
\caption{LPL Reaction rate constants in cm$^3$molec$^{-1}$s$^{-1}$ and half-lives ($\tau$) in days for the five gas phase association reactions.}             
\label{table:6}      
\centering          
  \begin{tabular}{cccc c|ccccccc}
  \hline
     &  iso-PrCN &&&& n-PrCN&&&& \\

      \hline
    
T(K)& C$_3$H$_7$+CN& $\tau$& CH$_3$+C$_2$H$_4$CN& $\tau$& C$_2$H$_5$+CH$_2$CN& $\tau$& CH$_3$+C$_2$H$_4$CN& $\tau$& CN+CH$_3$C$_3$H$_7$&$ \tau$ \\
     
      \hline

10&  9.450& 0.073& 2.740$\times 10^{-03}$& 2.5$\times 10^{+02}$& 1.010$\times 10^{-03}$& 6.9$\times 10^{+02}$& 2.150$\times 10^{-03}$& 3.2$\times 10^{+02}$& 1.120$\times 10^{+01}$& 0.062 \\
20& 3.880$\times 10^{-01}$& 1.8& 7.820$\times 10^{-05}$& 8.9$\times 10^{+03}$& 2.930$\times 10^{-05}$& 2.4$\times 10^{+04}$& 6.220$\times 10^{-05}$& 1.1$\times 10^{+04}$& 4.610$\times 10^{-1}$& 1.5 \\
30& 6.150$\times 10^{-02}$& 11& 9.700$\times 10^{-06}$& 7.1$\times 10^{+04}$& 3.800$\times 10^{-06}$& 1.8$\times 10^{+05}$& 8.030$\times 10^{-06}$& 8.6$\times 10^{+04}$& 7.320$\times 10^{-02}$& 9.5  \\
40& 1.690$\times 10^{-02}$& 41& 2.180$\times 10^{-06}$& 3.2$\times 10^{+05}$& 9.080$\times 10^{-07}$& 7.6$\times 10^{+05}$& 1.890$\times 10^{-06}$& 3.7$\times 10^{+05}$& 2.010$\times 10^{-02}$& 34  \\
50& 6.170$\times 10^{-03}$& 1.1$\times 10^{+02}$& 6.800$\times 10^{-07}$& 1.0$\times 10^{+06}$& 3.010$\times 10^{-07}$& 2.3$\times 10^{+06}$& 6.170$\times 10^{-07}$& 1.1$\times 10^{+06}$& 7.420$\times 10^{-03}$& 93 \\
60& 2.680$\times 10^{-03}$& 2.6$\times 10^{+02}$& 2.600$\times 10^{-07}$& 2.7$\times 10^{+06}$& 1.220$\times 10^{-07}$& 5.7$\times 10^{+06}$& 2.450$\times 10^{-07}$& 2.8$\times 10^{+06}$& 3.270$\times 10^{-03}$& 2.1$\times 10^{+02}$ \\
70& 1.310$\times 10^{-03}$& 5.3$\times 10^{+02}$& 1.140$\times 10^{-07}$& 6.1$\times 10^{+06}$& 5.660$\times 10^{-08}$& 1.2$\times 10^{+07}$& 1.110$\times 10^{-07}$& 6.2$\times 10^{+06}$& 1.630$\times 10^{-03}$& 4.3$\times 10^{+02}$  \\
80& 6.990$\times 10^{-04}$& 9.9$\times 10^{+02}$& 5.550$\times 10^{-08}$& 1.2$\times 10^{+07}$& 2.900$\times 10^{-08}$& 2.4$\times 10^{+07}$& 5.530$\times 10^{-08}$& 1.3$\times 10^{+07}$& 8.810$\times 10^{-04}$& 7.9$\times 10^{+02}$  \\
90& 3.960$\times 10^{-04}$& 1.8$\times 10^{+03}$& 2.920$\times 10^{-08}$& 2.4$\times 10^{+07}$& 1.600$\times 10^{-08}$& 4.3$\times 10^{+07}$& 2.970$\times 10^{-08}$& 2.3$\times 10^{+07}$& 5.090$\times 10^{-04}$& 1.4$\times 10^{+03}$  \\
100& 2.360$\times 10^{-04}$& 2.9$\times 10^{+03}$& 1.630$\times 10^{-08}$& 4.3$\times 10^{+07}$& 9.340$\times 10^{-09}$& 7.4$\times 10^{+07}$& 1.690$\times 10^{-08}$& 4.1$\times 10^{+07}$& 3.090$\times 10^{-04}$& 2.2$\times 10^{+03 }$ \\
110& 1.470$\times 10^{-04}$& 4.7$\times 10^{+03}$& 9.580$\times 10^{-09}$& 7.2$\times 10^{+07}$& 5.720$\times 10^{-09}$& 1.2$\times 10^{+08}$& 1.000$\times 10^{-08}$& 6.9$\times 10^{+07}$& 1.960$\times 10^{-04}$& 3.5$\times 10^{+03}$  \\
120& 9.410$\times 10^{-05}$& 7.4$\times 10^{+03}$& 5.860$\times 10^{-09}$& 1.2$\times 10^{+08}$& 3.640$\times 10^{-09}$& 1.9$\times 10^{+08}$& 6.190$\times 10^{-09}$& 1.1$\times 10^{+08}$& 1.280$\times 10^{-04}$& 5.4$\times 10^{+03}$  \\
130& 6.220$\times 10^{-05}$& 1.1$\times 10^{+04}$& 3.710$\times 10^{-09}$& 1.9$\times 10^{+08}$& 2.390$\times 10^{-09}$& 2.9$\times 10^{+08}$& 3.940$\times 10^{-09}$& 1.8$\times 10^{+08}$& 8.640$\times 10^{-05}$& 8.0$\times 10^{+03}$  \\
140& 4.210$\times 10^{-05}$& 1.6$\times 10^{+04}$& 2.410$\times 10^{-09}$& 2.9$\times 10^{+08}$& 1.620$\times 10^{-09}$& 4.3$\times 10^{+08}$& 2.580$\times 10^{-09}$& 2.7$\times 10^{+08}$& 5.960$\times 10^{-05}$& 1.2$\times 10^{+04}$  \\
150& 2.910$\times 10^{-05}$& 2.4$\times 10^{+04}$& 1.610$\times 10^{-09}$& 4.3$\times 10^{+08}$& 1.120$\times 10^{-09}$& 6.2$\times 10^{+08}$& 1.730$\times 10^{-09}$& 4.0$\times 10^{+08}$& 4.200$\times 10^{-05}$& 1.7$\times 10^{+04}$  \\
160& 2.050$\times 10^{-05}$& 3.4$\times 10^{+04}$& 1.100$\times 10^{-09}$& 6.3$\times 10^{+08}$& 7.890$\times 10^{-10}$& 8.8$\times 10^{+08}$& 1.180$\times 10^{-09}$& 5.9$\times 10^{+08}$& 3.010$\times 10^{-05}$& 2.3$\times 10^{+04}$  \\
170& 1.470$\times 10^{-05}$& 4.7$\times 10^{+04}$& 7.650$\times 10^{-10}$& 9.1$\times 10^{+08}$& 5.660$\times 10^{-10}$& 1.2$\times 10^{+09}$& 8.200$\times 10^{-10}$& 8.5$\times 10^{+08}$& 2.190$\times 10^{-05}$& 3.2$\times 10^{+04}$  \\
180& 1.070$\times 10^{-05}$& 6.5$\times 10^{+04}$& 5.410$\times 10^{-10}$& 1.3$\times 10^{+09}$& 4.130$\times 10^{-10}$& 1.7$\times 10^{+09}$& 5.800$\times 10^{-10}$& 1.2$\times 10^{+09}$& 1.620$\times 10^{-05}$& 4.3$\times 10^{+04}$  \\
190& 7.910$\times 10^{-06}$& 8.8$\times 10^{+04}$& 3.880$\times 10^{-10}$& 1.8$\times 10^{+09}$& 3.050$\times 10^{-10}$& 2.3$\times 10^{+09}$& 4.150$\times 10^{-10}$& 1.7$\times 10^{+09}$& 1.210$\times 10^{-05}$& 5.7$\times 10^{+04}$  \\
200& 5.910$\times 10^{-06}$& 1.2$\times 10^{+05}$& 2.820$\times 10^{-10}$& 2.5$\times 10^{+09}$& 2.280$\times 10^{-10}$& 3.0$\times 10^{+09}$& 3.010$\times 10^{-10}$& 2.3$\times 10^{+09}$& 9.180$\times 10^{-06}$& 7.6$\times 10^{+04}$  \\
210& 4.470$\times 10^{-06}$& 1.6$\times 10^{+05}$& 2.080$\times 10^{-10}$& 3.3$\times 10^{+09}$& 1.730$\times 10^{-10}$& 4.0$\times 10^{+09}$& 2.210$\times 10^{-10}$& 3.1$\times 10^{+09}$& 7.030$\times 10^{-06}$& 9.9$\times 10^{+04}$  \\
220& 3.410$\times 10^{-06}$& 2.0$\times 10^{+05}$& 1.550$\times 10^{-10}$& 4.5$\times 10^{+09}$& 1.320$\times 10^{-10}$& 5.3$\times 10^{+09}$& 1.640$\times 10^{-10}$& 4.2$\times 10^{+09}$& 5.430$\times 10^{-06}$& 1.3$\times 10^{+05}$  \\
230& 2.630$\times 10^{-06}$& 2.6$\times 10^{+05}$& 1.160$\times 10^{-10}$& 6.0$\times 10^{+09}$& 1.020$\times 10^{-10}$& 6.8$\times 10^{+09}$& 1.230$\times 10^{-10}$& 5.6$\times 10^{+09}$& 4.230$\times 10^{-06}$& 1.6$\times 10^{+05}$  \\
240& 2.040$\times 10^{-06}$& 3.4$\times 10^{+05}$& 8.830$\times 10^{-11}$& 7.8$\times 10^{+09}$& 7.900$\times 10^{-11}$& 8.8$\times 10^{+09}$& 9.260$\times 10^{-11}$& 7.5$\times 10^{+09}$& 3.320$\times 10^{-06}$& 2.1$\times 10^{+05}$  \\
250& 1.600$\times 10^{-06}$& 4.3$\times 10^{+05}$& 6.760$\times 10^{-11}$& 1.0$\times 10^{+10}$& 6.190$\times 10^{-11}$& 1.1$\times 10^{+10}$& 7.050$\times 10^{-11}$& 9.8$\times 10^{+09}$& 2.620$\times 10^{-06}$& 2.6$\times 10^{+05 }$ \\
260& 1.260$\times 10^{-06}$& 5.5$\times 10^{+05}$& 5.210$\times 10^{-11}$& 1.3$\times 10^{+10}$& 4.880$\times 10^{-11}$& 1.4$\times 10^{+10}$& 5.410$\times 10^{-11}$& 1.3$\times 10^{+10}$& 2.090$\times 10^{-06}$& 3.3$\times 10^{+05}$  \\
270& 1.000$\times 10^{-06}$& 6.9$\times 10^{+05}$& 4.050$\times 10^{-11}$& 1.7$\times 10^{+10}$& 3.870$\times 10^{-11}$& 1.8$\times 10^{+10}$& 4.170$\times 10^{-11}$& 1.7$\times 10^{+10}$& 1.670$\times 10^{-06}$& 4.2$\times 10^{+05}$  \\
280& 7.990$\times 10^{-07}$& 8.7$\times 10^{+05}$& 3.160$\times 10^{-11}$& 2.2$\times 10^{+10}$& 3.090$\times 10^{-11}$& 2.2$\times 10^{+10}$& 3.240$\times 10^{-11}$& 2.1$\times 10^{+10}$& 1.350$\times 10^{-06}$& 5.1$\times 10^{+05 }$ \\
290& 6.420$\times 10^{-07}$& 1.1$\times 10^{+06}$& 2.490$\times 10^{-11}$& 2.8$\times 10^{+10}$& 2.480$\times 10^{-11}$& 2.8$\times 10^{+10}$& 2.530$\times 10^{-11}$& 2.7$\times 10^{+10}$& 1.090$\times 10^{-06}$& 6.4$\times 10^{+05}$  \\
298& 5.390$\times 10^{-07}$& 1.3$\times 10^{+06}$& 2.050$\times 10^{-11}$& 3.4$\times 10^{+10}$& 2.080$\times 10^{-11}$& 3.3$\times 10^{+10}$& 2.080$\times 10^{-11}$& 3.3$\times 10^{+10}$& 9.190$\times 10^{-07}$& 7.5$\times 10^{+05}$   \\
300& 5.180$\times 10^{-07}$& 1.3$\times 10^{+06}$& 1.970$\times 10^{-11}$& 3.5$\times 10^{+10}$&  2.0$\times 10^{-11}$& 3.5$\times 10^{+10}$& 1.99$\times 10^{-11}$&3.5$\times 10^{+10}$ & 8.850$\times 10^{-07}$& 7.8$\times 10^{+05}$  \\
 \hline
 \end{tabular}
\end{table*}

\label{lastpage}

\end{document}